\documentclass[longauth]{aa}
\usepackage[varg]{txfonts}
\usepackage{graphicx}
\usepackage{natbib}
\bibpunct{(}{)}{;}{a}{}{,} 
\graphicspath{{immagini/}}
\usepackage{color}
\usepackage{siunitx}
\usepackage[normalem]{ulem}

\newcommand {\logRHK}{\ensuremath{\log R'_\mathrm{HK}}}
\usepackage{lipsum}
\title{The GAPS programme at TNG}
\subtitle{XXII. The GIARPS view of the extended helium atmosphere of HD\,189733\,b accounting for stellar activity \thanks{Based on observations made with the Italian {\it Telescopio Nazionale
			Galileo} (TNG) operated by the {\it Fundaci\'on Galileo Galilei} (FGG) of the
		{\it Istituto Nazionale di Astrofisica} (INAF) at the
		{\it  Observatorio del Roque de los Muchachos} (La Palma, Canary Islands, Spain).}}

\author{G.~Guilluy		\inst{\ref{unito},\ref{oato}} 
	\and V.~Andretta   \inst{\ref{capo}}
	\and F.~Borsa        \inst{\ref{Brera}}
	\and P.~Giacobbe	\inst{\ref{oato}}
	\and{A.~Sozzetti}		    \inst{\ref{oato}}
	\and E.~Covino		\inst{\ref{capo}}
	\and{V.~Bourrier}	    	\inst{\ref{Geneve}}
	\and{L.~Fossati}		    \inst{\ref{Fossati}}
	\and{A.~S.~Bonomo}	    	\inst{\ref{oato}}
	\and{M.~Esposito}		    \inst{\ref{esposito}}
	\and{M.~S.~Giampapa}		    \inst{\ref{Giampapa}}
	\and{A.~Harutyunyan}		\inst{\ref{tng}}
	\and{M.~Rainer}		        \inst{\ref{arcetri}}
	\and{M.~Brogi}            \inst{\ref{Brogi1},\ref{Brogi2},\ref{oato}}
	\and{G.~Bruno}		        \inst{\ref{catania}}
	\and{R.~Claudi}
	\inst{\ref{padova}}
	\and{G.~Frustagli}		     \inst{\ref{Brera},\ref{Frustagli}}
	\and{A.~F.~Lanza}		        \inst{\ref{catania}}
	\and{L.~Mancini}             \inst{\ref{mancini1},\ref{mancini2},\ref{oato}}
	\and{L.~Pino}		        \inst{\ref{Pino},\ref{arcetri}}
	\and{E.~Poretti}            \inst{\ref{Brera},\ref{tng}}
	\and{G.~Scandariato}		 \inst{\ref{catania}}
	\and{L.~Affer}
	\inst{\ref{palermo}}
	\and{C.~Baffa} \inst{\ref{arcetri}}
	\and{A.~Baruffolo} \inst{\ref{padova}}
	\and{S.~Benatti}
	\inst{\ref{palermo}}
	\and{K.~Biazzo}
	\inst{\ref{catania}}
	\and{A.~Bignamini}
	\inst{\ref{trieste}}
	\and{W.~Boschin}			\inst{\ref{tng},\ref{Boschin1},\ref{Boschin2}}
	\and{I.~Carleo}
	\inst{\ref{carleo}}
	\and{M.~Cecconi} \inst{\ref{tng}}
	\and{R.~Cosentino}              \inst{\ref{tng}}
	\and{M.~Damasso}  
	\inst{\ref{oato}}
	\and{M.~Desidera}  
	\inst{\ref{padova}}
	\and{G.~Falcini}\inst{\ref{arcetri}}		\and{A.~F.~Martinez~Fiorenzano}             \inst{\ref{tng}}
	\and{A.~Ghedina} \inst{\ref{tng}}
	\and{E.~González-Álvarez} \inst{\ref{gonzalez}}
	\and{J.~Guerra} \inst{\ref{tng}}
	\and{N.~Hernandez} \inst{\ref{tng}}
	\and{G.~Leto}		        \inst{\ref{catania}}
	\and{A.~Maggio}		        \inst{\ref{palermo}}
	\and{L.~Malavolta}		        \inst{\ref{catania}}
	\and{J.~Maldonado}		        \inst{\ref{palermo}}
	\and{G.~Micela} \inst{\ref{palermo}}
	\and{E.~Molinari} \inst{\ref{Cagliari}}
	\and{V.~Nascimbeni}  \inst{\ref{padova}}
	\and{I.~Pagano}  \inst{\ref{catania}}
	\and{M.~Pedani}             \inst{\ref{tng}}
	\and{G.~Piotto}             \inst{\ref{piotto}}
	\and{A.~Reiners}              \inst{\ref{reiners}}
}

\institute{Dipartimento di Fisica, Universit\`{a} degli Studi di Torino, via Pietro Giuria 1, I-10125 Torino, Italy \label{unito}
	\and INAF — Osservatorio Astronomico di Capodimonte, Salita Moiariello 16, I-80131 Naples, Italy \label{capo}
	\and INAF – Osservatorio Astrofisico di Torino, Via Osservatorio 20, I-10025 Pino Torinese, Italy \label{oato}
	\and INAF – Osservatorio Astronomico di Brera, Via E. Bianchi 46, 23807 Merate (LC), Italy \label{Brera}
	\and Observatoire de l'Universit\'e de Gen\`eve, 51 chemin des Maillettes, 1290 Versoix, Switzerland \label{Geneve}
	\and Space Research Institute, Austrian Academy of Sciences, Schmiedlstrasse 6, 8042 Graz, Austria \label{Fossati}
	\and Thüringer Landessternwarte, Tautenburg Sternwarte 5 - 07778 Tautenburg, Germany \label{esposito}
	\and National Solar Observatory, Tucson, AZ 85719, USA and Steward Observatory, University of Arizona, Tucson, AZ  85721 \label{Giampapa}
	\and Fundaci\'on G. Galilei - INAF (Telescopio Nazionale Galileo), Rambla J. A. Fern\'andez P\'erez 7, E-38712 Bre\~na Baja (La Palma), Spain \label{tng}
	\and INAF – Osservatorio Astrofisico di Arcetri, Largo E. Fermi 5, 50125 Firenze, Italy \label{arcetri}
	\and Department of Physics, University of Warwick, Gibbet Hill Road, Coventry, CV4 7AL, UK \label{Brogi1}
	\and Centre for Exoplanets and Habitability, University of Warwick, Gibbet Hill Road, Coventry, CV4 7AL, UK \label{Brogi2}
	\and INAF – Osservatorio Astrofisico di Catania, Via S. Sofia 78, 95123,Catania, Italy \label{catania}
	\and INAF – Osservatorio Astronomico di Padova, Vicolo dell’Osservatorio 5, 35122, Padova, Italy \label{padova}
	\and Dipartimento di Fisica G. Occhialini, Università degli Studi di Milano-Bicocca, Piazza della Scienza 3, 20126 Milano, Italy \label{Frustagli}
	\and Department of Physics, University of Rome Tor Vergata, Via della Ricerca Scientifica 1, 00133 Rome, Italy \label{mancini1}
	\and Max Planck Institute for Astronomy, K$\rm{\ddot{o}}$nigstuhl 17, 69117 Heidelberg, Germany \label{mancini2}
	\and Anton Pannekoek Institute for Astronomy, University of Amsterdam, Science Park 904, 1098 XH Amsterdam, The Netherlands \label{Pino} 
	\and INAF – Osservatorio Astronomico di Palermo, Piazza del Parlamento, 1, 90134, Palermo, Italy \label{palermo}
	\and INAF – Osservatorio Astronomico di Trieste, via Tiepolo 11, 34143 Trieste, Italy \label{trieste}
	\and Instituto de Astrof\'{\i}sica de Canarias, C/V\'{\i}a L\'actea s/n, E-38205 La Laguna (Tenerife), Spain \label{Boschin1} 
	\and Departamento de Astrof\'{\i}sica, Univ. de La Laguna, Av. del Astrof\'{\i}sico Francisco S\'anchez s/n, E-38205 La Laguna (Tenerife), Spain \label{Boschin2} 
	\and Astronomy Department, 96 Foss Hill Drive, Van Vleck Observatory 101, Wesleyan University, Middletown, CT 06459, US \label{carleo}
	\and  Centro de Astrobiología (CSIC-INTA), Carretera de Ajalvir km 4 - 28850 Torrejón de Ardoz, Madrid, Spain \label{gonzalez}
	\and INAF – Osservatorio di Cagliari, via della Scienza 5, I-09047 Selargius, CA, Italy \label{Cagliari}
	\and Dip. di Fisica e Astronomia Galileo Galilei – Universit`a di Padova, Vicolo dell’Osservatorio 2, 35122, Padova, Italy \label{piotto}
	\and Institut f$\rm{\ddot{u}}$r Astrophysik, Friedrich-Hund Platz 1, D-37077 G$\rm{\ddot{o}}$ttingen, Germany \label{reiners}
}

\date{Received <date> / Accepted <date>}

\abstract{Exoplanets orbiting very close to their parent star are strongly irradiated. This can lead the upper atmospheric layers to expand and evaporate into space.
	The metastable helium (\ion{He}{I}) triplet at 1083.3~nm has recently been shown to be a powerful diagnostic to probe extended and escaping exoplanetary atmospheres.}
{We perform high-resolution transmission spectroscopy of the transiting hot Jupiter HD\,189733\,b with the GIARPS (GIANO-B + HARPS-N) observing mode of the Telescopio Nazionale Galileo, taking advantage of the simultaneous optical+near infrared spectral coverage to detect \ion{He}{I} in the planet's extended atmosphere and to gauge the impact of stellar magnetic activity on the planetary absorption signal. Observations were performed during five transit events of HD\,189733\,b.}
{By comparison of the in- and out-of-transit GIANO-B observations we compute high-resolution transmission spectra. We then utilize them to perform equivalent width measurements and carry out  light-curves analyses in order to consistently gauge the excess in-transit absorption in correspondence of the \ion{He}{I} triplet.}
{We spectrally resolve the \ion{He}{I} triplet and detect an absorption signal during all five transits. The mean in-transit absorption depth amounts to $0.75 \pm 0.03$\,$\%$\,(25$\sigma$) in the core of the strongest helium triplet component. We detect night-to-night variations in the \ion{He}{I} absorption signal likely due to the transit events occurring in presence of stellar surface inhomogeneities. We evaluate the impact of stellar-activity pseudo-signals on the true planetary absorption using a comparative analysis of the \ion{He}{I}~1083.3~nm (in the near-infrared, nIR) and the H$\alpha$ (in the visible) lines. We interpret the time-series of the \ion{He}{I} absorption lines in the three nights not affected by stellar contamination -exhibiting a mean in-transit absorption depth of 0.77 $\pm$ 0.04 $\%$ (19$\sigma$) in full agreement with the one derived from the full dataset- using a 3-d atmospheric code. In agreement with previous results, our simulations suggest that the helium layers only fill part of the Roche lobe. Observations can be explained with a thermosphere heated to $\sim$12000\,K, expanding up to $\sim$ 1.2 planetary radii, and losing $\sim1$~g\,s$^{-1}$ of metastable helium.}{Our results reinforce the importance of simultaneous optical+nIR monitoring when performing high-resolution transmission spectroscopy of hot planets' extended and escaping atmospheres in the presence of stellar activity.}

\titlerunning{The HD\,189733\,b extended helium atmosphere with GIARPS}
\authorrunning{G.~Guilluy et al.}
\begin{document}  
	\maketitle
	
	\section{Introduction}\label{sec:intro}
	
	Exoplanets that orbit very close to their host stars are subject to extreme physical processes (e.g. hydrodynamic escape). For instance, the intense X-ray and extreme ultraviolet (EUV) irradiation received from the parent stars can lead the gas content in the upper layers of their atmosphere to expand and reach high velocities. The gas fraction with a velocity greater than the escape velocity can then overcome the planetary force of gravity, escaping into space.  
	Atmospheric evaporation processes driving strong mass-loss might be responsible for the paucity of intermediate mass/size (sub-Jovian) planets with periods $\le$3 d (the so-called `Neptunian desert'. See e.g. \citealt{Lecavelier2007,Beaug2013,Lundkvist2016, Mazeh2016, Owen2018}) observed in the mass(radius)-period distribution of close-in exoplanets.

	Hydrogen has long been investigated as a tracer for atmospheric evaporation because it is the lightest and most-abundant element in giant planets, and therefore it can easily escape, and produce an extended exosphere surrounding the planet. The first observation of this effect was obtained for the transiting hot Jupiter HD\,209458\,b by \citet{Vidal-Madjar2003}. Based on space-borne HST/STIS spectroscopy,  strong variations in the ultraviolet Ly-$\alpha$ emission line at 121.567~nm between in-transit and out-of-transit spectra demonstrated the presence of an extended comet-like tail of escaping hydrogen atoms from HD\,209458\,b.  Additional evidence for atmospheric escape in other giants has been obtained using Ly-$\alpha$ spectroscopy \citep[e.g.,][]{Lecavelier2010, Lecavelier2012, Bourrier2013, Kulow2014,Ehrenreich2015,Levie2017,Bourrier2018}.

	However, this proxy is strongly affected by interstellar absorption and by geocoronal emission (due to the Earth's exosphere).  Therefore, only the wings of the Ly-$\alpha$ line can be used to probe escaping exoplanetary atmospheres \citep[e.g.,][]{Ehrenreich2015}.\\

	The 1083.3~nm \ion{He}{I} near-infrared (nIR) triplet (vacuum wavelength)\footnote{In this paper we use vacuum wavelengths for GIANO-B and air wavelengths for HARPS-N. This is due to the type of reference system selected as default by the reduction software used.} is very weakly affected by interstellar absorption and it can be observed from the ground. It is a well-known diagnostic used in a variety of astrophysical contexts, for instance to study the structure of the solar chromosphere and transition region, prominences, flares \citep[e.g.,][]{Andretta2008}, magnetic activity in cool stars \citep[e.g.,][]{Zarro1986}, the dynamics of stellar winds \citep[e.g.,][]{Dupree1992}, the outflows from quasars \citep[e.g.,][]{Leighly2011}, and to trace planetary nebulae \citep{Weidmann2013}. The \ion{He}{I} nIR triplet was first suggested to be a relevant absorption signature in the spectra of exoplanetary atmospheres by \citet{Seager2000}, while more recent theoretical work \citep{Oklop2018} revisited those findings to infer good prospects for the observability of extended atmospheres of known hot planets through this channel. 
	
	Prompted by the theoretical expectations, early attempts to detect the presence of helium in the extended atmospheres of exoplanets employing medium-resolution spectroscopy \citep{Moutou2003} only succeeded in placing upper limits on this absorption feature. However, the tide has recently turned, thanks to renewed efforts carried out both at low and high spectral resolution in space and from the ground, respectively. Using the Wide Field Camera 3 (WFC3) on board HST, an excess absorption in the helium triplet was detected in the atmosphere of the two warm Neptunes WASP-107b and HAT-P-11b \citep{Spake2018, Mansfield2018}. Using CARMENES (Calar Alto high-Resolution search for M dwarf with Exoearths with Near-infrared and optical \'Echelle Spectrographs) data, \citet{Nortmann2018}, \citet{Allart2018}, \citet{Salz2018}, \citet{Allart2019}, \citet{Alonso2019b} confirmed the presence of extended atmospheres of helium in HAT-P-11b and WASP-107b, and reported new detections of \ion{He}{I} absorption in HD\,189733\,b, WASP-69b, and HD\,209458\,b.
	
	However, all excited levels of helium, including the lower level of the triplet 1083.3~nm line, the metastable level $1s2s\: ^3\!S$, lie almost 20 eV above the $1s^2\: ^1\!S$ ground state. In plasmas at temperatures below a few $10^4$~K, the level population of that state, and of all the triplet states, is usually negligible. Indeed, no \ion{He}{I} line is produced in the photosphere of stars of spectral type later than A, which includes the majority of stars hosting hot exoplanets. 
	Thus, in contrast to the Ly-$\alpha$ diagnostic of atmospheric evaporation, which arises from the ground level of \ion{H}{I}, some non-thermal process is required to populate the levels producing the observed absorption in the \ion{He}{I}~1083.3~nm triplet. 
	
	A consensus has not emerged yet on the nature of the non-thermal processes responsible for the observed \ion{He}{I}~1083.3~nm line strength in the outer atmosphere of the Sun and of solar-type stars.  Various non-equilibrium collisional processes have been proposed \citep[e.g.,][]{Andretta2000,MacPherson1999,Golding2016}. As an alternative, a connection with the solar and stellar coronal emission has often been invoked.
	It had indeed been very early pointed out \citep{Goldberg1939,Hirayama1971} that EUV radiation below 50.4~nm could photoionize neutral He atoms.  
	Subsequent recombination cascades can then populate the triplet system, and in particular the $1s2s ^3S$ level. The depopulation of this triplet state (which can only radiatively decay through a forbidden-line photon) progresses slowly, 10.9 day$^{-1}$ \citep{Drake1971}, making it metastable. Atoms in that state can therefore efficiently scatter nIR photons at 1083.3~nm. The process has been described by \cite{Zirin1975} in the context, for example, of solar and stellar chromospheres, and analysed in detail by \cite{Andretta1997}, but it can also act in exoplanet escaping atmospheres where it could produce the strong absorption feature in the transmission spectrum \citep{Seager2000}.
	
	It is therefore clear that, in absence of sufficient EUV illumination from the host star, no absorption features from excited levels of \ion{He}{I} in the transmission spectrum of exoplanet escaping atmospheres can be observed. Since in solar-type or cool stars, such an EUV radiation is only associated with stellar activity phenomena, we expect signatures of exoplanet escaping atmospheres to be clearly observable in the \ion{He}{I}~1083.3~nm line only in exoplanets orbiting active or moderately active stars \citep{Oklop2019}. This implies that, for a correct quantitative interpretation of observed absorption in the \ion{He}{I} 1083.3~nm line, it is imperative to obtain simultaneous estimates of the activity level of the host star.
	
	The presence of inhomogeneities on the stellar surface can also complicate the analysis of transmission spectra in a different respect: the planetary disk transiting for example over quiet stellar regions, i.e. regions with below-average \ion{He}{I} absorption, can mimic a pseudo-absorption at the position of the helium triplet.  Conversely, the occultation of active regions by the transiting exoplanet could produce an emission signal in the transmission spectrum which could reduce the absorption signal related to the planetary atmosphere. In their study focused on HD\,189733\,b, \citet{Salz2018} discussed how these pseudo-signals, due to stellar activity, can interfere with the atmospheric absorption signal, but without a clear evaluation of these effects and how they can be distinguished. However, the analogy with the Sun suggests that \ion{He}{I} 1083.3~nm line in active, solar-like stars is much more sensitive to the presence of plage-like regions or stellar prominences than to starspots. This raises the possibility that a better understanding of the relative contribution to observed helium absorption features could be gleaned by combined analyses of the \ion{He}{I} line and other activity-sensitive spectral diagnostics. 
	
	In this paper we present a new investigation of the extended atmosphere of the Hot Jupiter HD\,189733\,b at high spectral resolution both in the optical and nIR using simultaneous observations gathered with the GIANO-B and HARPS-N spectrographs (\S~\ref{sec:obs}). We employ a multi-technique approach to confirm the recent detection of helium in the planet's extended atmosphere (\S~\ref{Sect3}), and describe a new methodology to consistently evaluate the impact of stellar activity pseudo-signals on the true absorption by circumplanetary material (\S~\ref{sec:discussion}). We then interpret the helium absorption observations not significantly affected by stellar contamination in terms of effective mass-loss, using detailed 3-d simulations of the atmosphere of HD\,189733\,b (\S~\ref{EVE}). Finally, we summarize our results and conclude (\S~\ref{sec:conclusions}) by highlighting possible future applications of this method to put more constraints on the real planetary absorption.

	\section{Observations}\label{sec:obs}
	
\begin{figure}
		\centering
		\includegraphics[width=9cm]{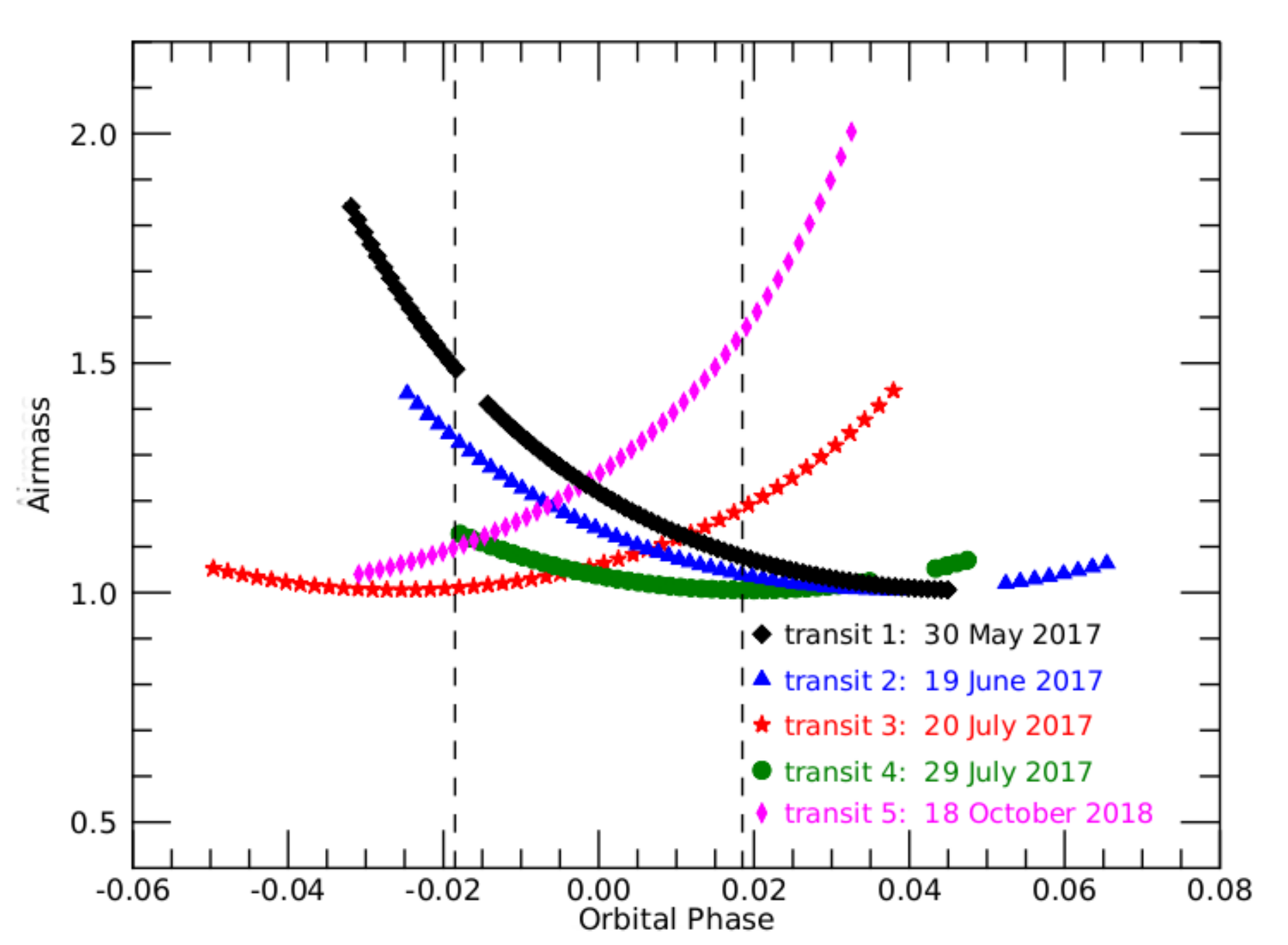}
		\caption{Airmass during the GIARPS observations. The colour-symbol coding presented here will be adopted consistently throughout the work. The contact points $t_1$ and $t_4$ are marked with vertical dashed lines.}
		\label{airmass_}
	\end{figure}

	We observed the system HD\,189733 using the GIARPS observing mode of the Telescopio Nazionale Galileo (TNG) telescope \citep{Claudi2017}, which, by obtaining high-resolution spectra with the HARPS-N (resolving power $R\sim$ 115,000) and GIANO-B ($R\sim$ 50,000) spectrographs, allows for simultaneous coverage over the optical (0.39-0.69\,$\mu$m for HARPS-N) and nIR ($\&$ 0.95-2.45\,$\mu$m for GIANO-B) wavelength ranges. The full dataset encompasses a total of five primary transit events of HD\,189733\,b, three observed on UT 30 May 2017, UT 20 July 2017, and UT 18 October 2018 within the context of the GAPS Project, and two scheduled on UT 19 June 2017 and UT 9 July 2017 as part of programme AOT35\_14 (P.I.: V. Andretta). A log of the observations with the number of collected spectra, exposure times, and achieved signal-to-noise ratio S/N is given in Table \ref{table:1}. The target was observed at airmass as low as 1.005 and as high as 2.004 (Fig.~\ref{airmass_}). Table \ref{table:2} lists the system parameters adopted in this work.
	\begin{table*}
		\caption{HD\,189733\,b GIARPS observations log.}
		\label{table:1} 
		\small
		\centering         
		\begin{tabular}{c | c | c | c c| c c| c c}          
			\hline\hline                       
			Night &  Transit number & Programme & \multicolumn{2}{c|}{N$_{\rm obs}$} & \multicolumn{2}{c|}{Exposure time} & \multicolumn{2}{c}{S/N$_{\rm{AVE}}$\tablefootmark{a}} \\
			
			& & & HARPS-N & GIANO-B & HARPS-N & GIANO-B & HARPS-N & GIANO-B \\
			\hline
			30 May 2017 & 1 & GAPS & 46 & 88 & 300s & 100s & 133 & 58 \\
			19 June 2017 & 2 & A35TAC\_14 & 44 & 56 & 300s & 200s & 106 & 116\\
			20 July 2017 & 3 & GAPS & 55 & 48 & 300s & 300s & 147 & 131 \\
			29 July 2017 & 4 & A35TAC\_14 & 40 & 44 & 300s & 200s  & 129 & 101 \\
			18 October 2018 & 5 & GAPS & 15& 48 & 900s & 200s & 186 & 76 \\

			\hline 
		\end{tabular}
		\tablefoot{
			\tablefoottext{a}{Time-averaged S/N of the order containing the \ion{He}{I} triplet (for GIANO-B spectra) and the H$\alpha$ line (for HARPS-N spectra). }
		}
	\end{table*}

	\begin{table}
		\caption{Stellar and planetary parameters adopted in this work.}             
		\label{table:2}     
		\centering  
		\begin{tabular}{l c l}          
			\hline\hline                       
			Parameters &  Value & Reference \\ 
			\hline 
			\multicolumn{2}{l}{\textbf{Planetary and transit parameters}} & \\
			K$_{\rm P}$ & $152.5_{-1.8}^{+1.3}$ km s$^{-1}$ &\citet{Brogi2018}\\
			T$_{\rm 0}$ [BJD$_{\rm UTC}]$ & $2454279.436714(15)$  &\citet{Agol2010} \\
			P & $2.21857567(15)$ d &\citet{Agol2010}\\
			i & $85.710(24)$ deg &\citet{Agol2010} \\
			b & $0.6631(23)$  &\citet{Agol2010} \\
			$\rho_{\rm P}$ & 0.943(24) g cm$^{-3}$  &\citet{Agol2010} \\
			R$_{\rm P}$/R$_\star$ &  $0.155313(188) $ &\citet{Agol2010} \\
			a/R$_\star$  & $8.863(20)$ &\citet{Agol2010}\\
			\hline
			\multicolumn{2}{l}{\textbf{Stellar parameters}} & \\
			K$_{\rm s}$ & $201.96_{-0.63}^{+1.07}$ m s$^{-1}$ &\citet{Triaud2009}\\
			V$_{\rm sys}$ & $-2.361(3)$ km s$^{-1}$ &\citet{Bouchy2005}\\
			B-V & 0.930 & \citet{Koen2010} \\
			\hline                                       
		\end{tabular}
	\end{table}
	
	\medskip

	\subsection{GIANO-B Data} \label{giano_}
	
	In order to be operated in GIARPS mode, the nIR high-resolution spectrograph GIANO \citep{Oliva2006} was recently moved at the Nasmyth-B focal station of the TNG \citep{Carleo2018}, and re-named GIANO-B. The GIANO-B echellogram has a fixed format and includes 50 orders. The spectra are imaged on a HAWAII-2 2048 $\times$ 2048 detector. The data consist of a sequence of nodded observations, with the target observed at predefined A and B positions on the slit, following an ABAB pattern  \citep{Claudi2017}. For each nodding sequence, the thermal background and telluric emission lines are  monitored at the slit position that is not illuminated by the target, and subsequently subtracted. 
	
	The GIANO-B spectra are extracted, blaze corrected, and wavelength calibrated using the GOFIO data reduction pipeline \citep{Rainer2018}.\\

	\subsection{HARPS-N Data}
	
	HARPS-N is the high-resolution, fibre-fed, cross-dispersed echelle spectrograph mounted at the Nasmyth-B focus of the TNG \citep{Cosentino2012}. The observations were carried out using the \verb+objAB+ observational setup, with fibre A on the target and fibre B on the sky. The fibre entrance is reimaged onto a 4k $\times$ 4k CCD, where echelle spectra of 69 orders are formed for each fibre. 
	The HARPS-N data are reduced with the standard Data Reduction Software (DRS), version 3.7 \citep[DRS,][]{Cosentino2012}. 
	
	\section{Data analysis of the \ion{He}{I}~1083.3~nm triplet} \label{Sect3}
	
	\subsection{Further treatment of GIANO-B data}	
	
	For the purpose of our scientific analysis, the GIANO-B spectra reduced by the GOFIO pipeline require additional processing steps, as follows: 
	\begin{enumerate}
		\item \emph{Wavelength calibration refinement}
		\\	Temporal variations in the wavelength solution due to a non-optimal stability of the GIANO-B spectrograph are removed using the approach described in \citet{Brogi2018}, which is based on a spline interpolation procedure. 	The initial wavelength calibration, obtained using an U-Ne lamp, is then refined employing a telluric spectrum as in \citet{Brogi2018}. The magnitude of these wavelength calibration refinements, for the considered nights, and in the region around the \ion{He}{I} triplet, is $\sim$1.3~km s$^{-1}$, lower than an individual GIANO-B resolution element (1 pixel = 2.7~km s$^{-1}$)
		\\
		\item \emph{Telluric lines removal}
		
		When working with ground-based spectra, we must take into account that they are contaminated by the imprints of the Earth's atmosphere. Therefore, our data must be corrected from telluric features. 
		First of all, for each night, we normalize the spectra by dividing them by the average flux computed in two intervals on the blue (1082.6-1082.8~nm)  and red (1083.9-1084.0~nm) sides of the \ion{He}{I} triplet lines. To take into account possible variations between the two nodding positions, the spectra acquired in the nodding position A (A-spectra) are treated separately from those acquired in the position B (B-spectra). In order to recognize and remove the telluric contamination, we follow the methods implemented by \citet{Snellen2008, Vidal-Madjar2010} and \citet{Astudillo-Defru2013}. These techniques consider that the logarithm of the telluric lines' strength increases linearly with the airmass. This is a consequence of the solution of the radiative transfer equation assuming only absorption.
		The total intensity at a certain wavelength $I{_{\rm{0\lambda}}}$, reaching the top of the atmosphere, is converted into an observed spectrum $I{_{\rm{\lambda}}}$ according to the following formula \citep{Vidal-Madjar2010}: 
		\begin{equation}
			I{_{\rm{\lambda}}}=T(\lambda)^a \times I{_{\rm{0\lambda}}},
			\label{RTE}
		\end{equation}
		where $T(\lambda)$ is the vertical atmospheric transmittance, i.e. the telluric spectrum, and $a$ is the airmass. For each night and nodding position, we derive $T(\lambda)$, modeling the relationship in Eq.~(\ref{RTE}) via linear regression \citep{Wyttenbach2015}. This operation is performed only using the out-of-transit spectra because, during transit, the presence of the planetary atmosphere could leave detectable imprints in $I_{\lambda}$. The only exception is transit 2 for which, due to the lack of a sufficient number of observations before the transit, we have to use the in-transit spectra too to compute a high-quality
		telluric spectrum \citep[see, e.g.,][]{Wyttenbach2015}.
		In Fig.~\ref{spettro_tell_} we show the telluric spectrum (in black) built for the night with the highest S/N (20 July 2017) considering only the A-spectra. This is consistent with the telluric spectrum generated via the ESO Sky Model Calculator  \footnote{\url{https://www.eso.org/observing/etc/bin/gen/form?INS.MODE=swspectr+INS.NAME=SKYCALC}} shown in blue.
		
		\noindent{Each stellar spectrum is then corrected from the telluric contamination by dividing it by $T(\lambda)^a$.} 
		We note that some strong telluric features are not fully removed. 
		This might be due to either short timescale variations in telluric species content or second-order deviations from the linear dependence with the airmass \citep{Astudillo-Defru2013}. However, since these telluric residuals do not fall in the wavelength ranges used in the presented analysis, our results are not affected.
		\\
		
		Generally, observations from the ground are also contaminated by telluric emission lines. In particular, in the spectral region of our interest, there are three OH emission lines that fall near the \ion{He}{I} triplet (more precisely at $\lambda~1083.21$~nm, $\lambda~1083.24$~nm, and $\lambda~1083.43$~nm). However, as the GIANO-B observations are acquired with a nodding pattern that allows for thermal background and emission lines subtraction at the level of the standard data extraction pipeline (see \S~\ref{giano_}), at this stage of the analysis we do not need to manually correct for the OH lines anymore, as it is instead done in other works  \citep[e.g.][]{Salz2018, Allart2019, Nortmann2018}. 
		\\	
		\item \emph{Fringing correction}

		The telluric spectrum in Fig.~\ref{spettro_tell_} clearly shows a sinusoidal fringing pattern, which is known to affect high S/N GIANO-B spectra in the central/red part of the blue orders\footnote{\url{https://atreides.tng.iac.es/monica.rainer/gofio/blob/master/GOFIO_manual.pdf} \label{footnote_1}}. The effect is caused by the sapphire substrate ($\sim$0.38\,mm thick) placed above the sensitive part of the detector that, behaving like a Fabry-P\'erot, generates interference fringing with peak-to-peak distance approximately equal to $0.75\;(wl/1000)^2$\;nm, where $wl$ is the wavelength expressed in nm. Such fringing patterns must be corrected for if we want to study the \ion{He}{I} triplet. In this work, we mitigate fringing effects adopting two different approaches, one implemented at the level of the original spectra (\S~\ref{stellarvariability}), and the second when working with the transmission spectra (\S~\ref{tra_spec}, \ref{EW_par}). In Appendix~\ref{App_A} we discuss the impact of these two techniques on the determination of the \ion{He}{I} absorption levels.
		
		\begin{figure}
			\resizebox{\hsize}{!}{\includegraphics{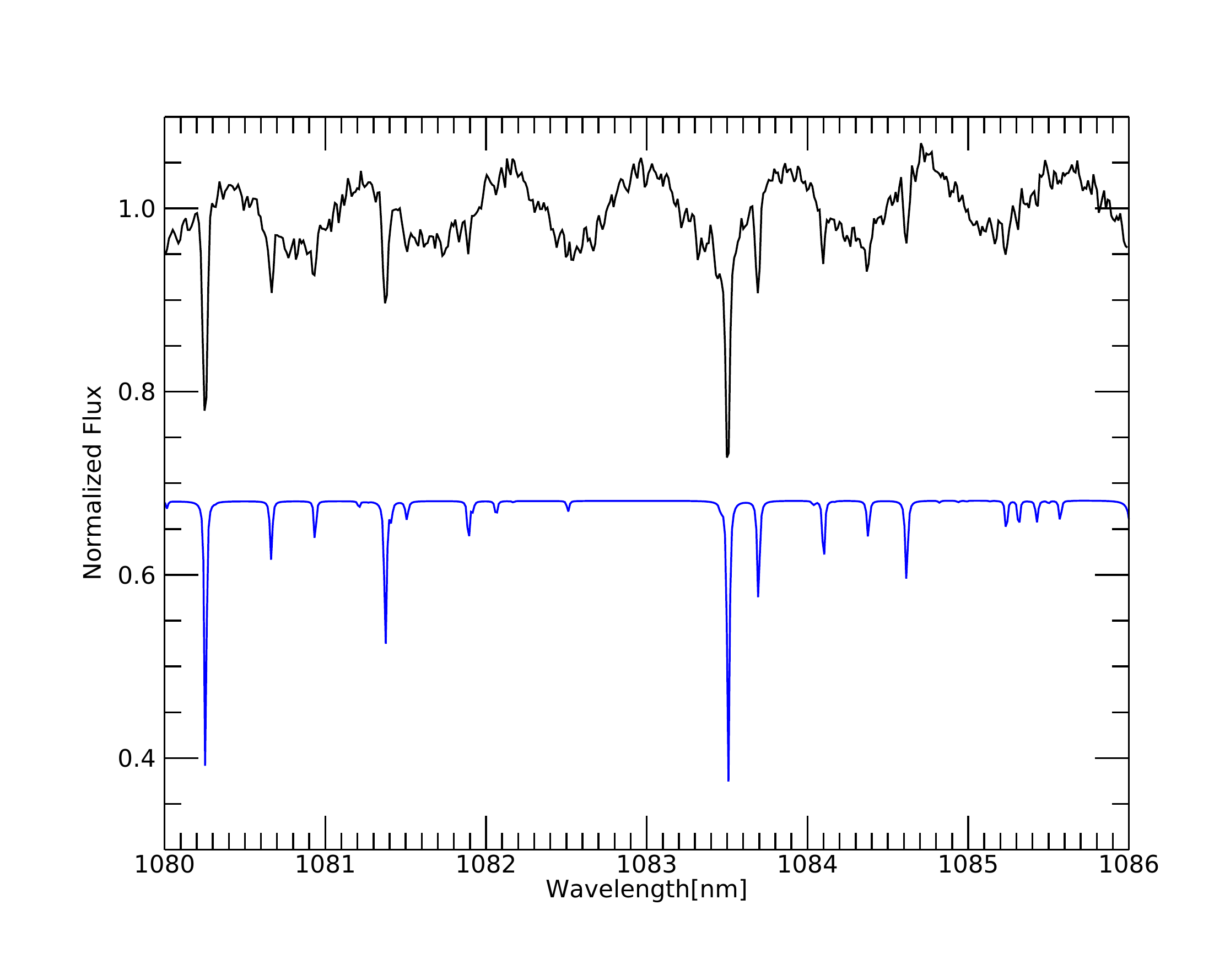}}
			\caption{Telluric absorptions spectrum built for the highest S/N night considering only the GIANO-B A-spectra. As a comparison, the telluric spectrum generated via the ESO SKy Model Calculator is shown in blue with an offset of -0.3. The two spectra are consistent. In our telluric spectrum the fringing pattern affecting GIANO-B spectra is also visible.}
			\label{spettro_tell_}
		\end{figure}
	\end{enumerate}
	\medskip
	The GIANO-B spectra subjected to wavelength calibration refinement and telluric line removal are then used to investigate the extended atmosphere of HD\,189733\,b, searching for planetary helium absorption in correspondence of the 1083.3~nm triplet with a multi-technique approach: \\
	\\
	\renewcommand{\labelitemi}{$\bullet$}
	\begin{itemize}
		\item Transmission spectroscopy and tomography
		\item Transmission spectra equivalent widths 
		\item Light-curve analysis
	\end{itemize}
	
	\subsection{Transmission spectroscopy and tomography}\label{tra_spec}
	
	When the planet passes in front of its host star, a transmission spectrum can be measured, which represents the the fractional area of the stellar disk covered by the planetary atmosphere as a function of wavelength, and used to investigate the optically thin regions of the planetary atmosphere. \\
	We compute transmission spectra as follows. The telluric-corrected GIANO-B spectra are initially shifted into the stellar rest frame by accounting for the barycentric Earth radial velocity, the stellar reflex motion induced by the planet, and the systemic velocity. Then, by averaging the out-of-transit A-spectra (B-spectra), a master-out spectrum $M_{\rm A}$ ($M_{\rm B}$) is created. All the A-spectra (B-spectra) are successively divided by $M_{\rm A}$ ($M_{\rm B}$) to create the Transmission Spectra $T_{\rm A}$ ($T_{\rm B}$), which are then corrected for fringing effects using Method \#1 described in Appendix~\ref{App_A}. 
	
	The two panels of Fig.~\ref{tom_he_star} show the two-dimensional map of the full time series of fringing-corrected transmission spectra of HD\,189733\,b in wavelength-orbital phase space in the neighborhood of the \ion{He}{I} triplet lines. The tomographic representation is routinely used to highlight the true reference frame of each absorption/emission signal \citep[e.g.,][]{Borsa2018}. When applying the tomography in the stellar reference frame (right panel of Fig.~\ref{tom_he_star}), an excess of absorption is clearly present during the transit at the position of the stellar helium lines. The signal follows the planetary radial velocity, revealing its planetary origin\footnote{Given the negligible impact of the Rossiter–McLaughlin effect (RME) during transit on the analysis and interpretation of the planetary helium absorption signal \citep[see \S~\ref{EVE} and][]{Salz2018}, we do not correct for it.}.
	\begin{figure*}
		\centering
		\includegraphics[width=\linewidth]{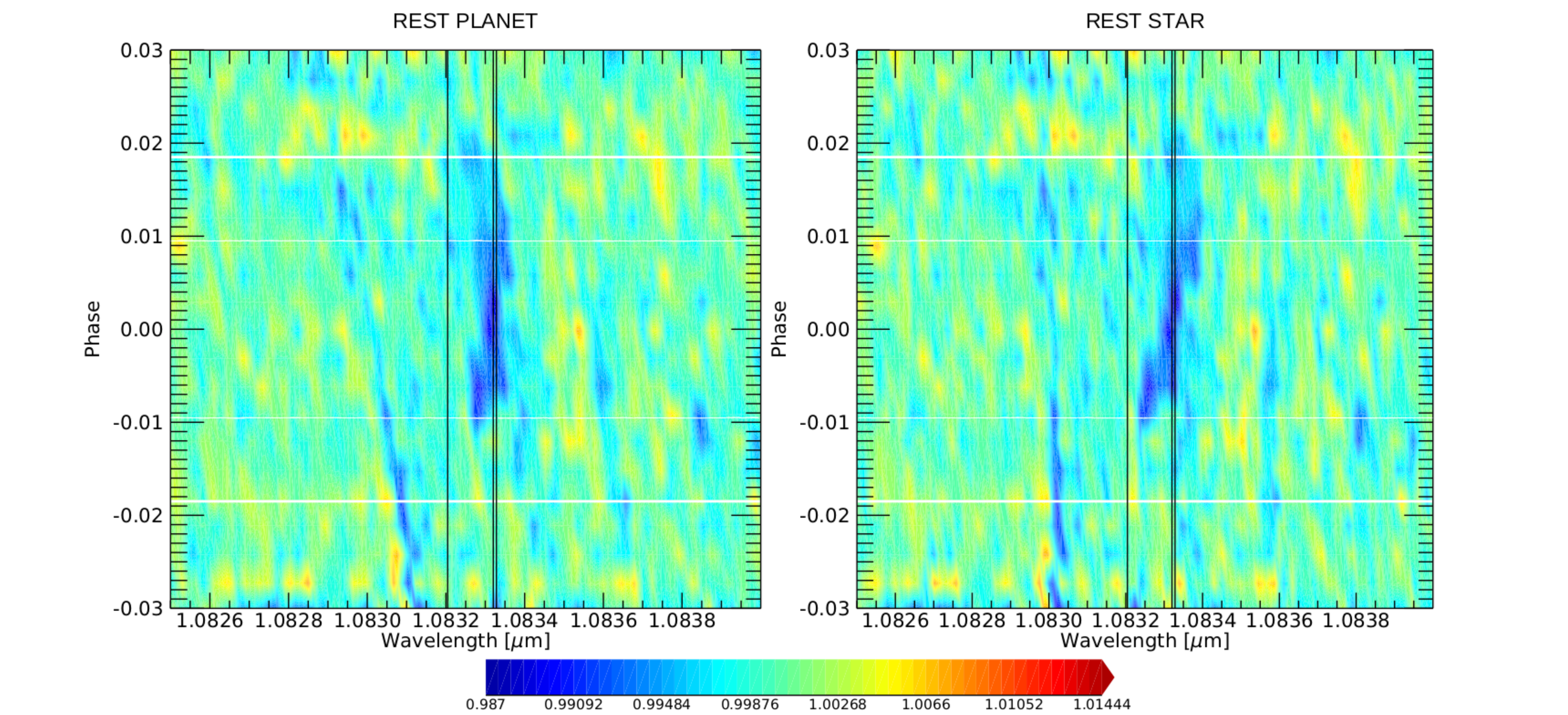}
		\caption{Transmission spectra shown in tomography in the planetary (left) and stellar (right)  rest frame, as a function of the wavelength and the orbital phase (transmission spectra are binned both in wavelength and in phase). An excess absorption (in blue) is present during the transit, the three helium triplet lines are indicated with vertical black lines. The contact point $t_1$, $t_2$, $t_3$ and $t_4$ are marked with horizontal white lines.}
		\label{tom_he_star}
	\end{figure*} 
	At this point, we shift the transmission spectra in the planetary rest frame (left panel of Fig.~\ref{tom_he_star}) and, for the two nodding positions, we build a mean transmission spectrum as the average of the $T_{\rm A}$ ($T_{\rm B}$) between the second ($t_2$) and the third ($t_3$) contact points. Then, for every night, by averaging these two mean transmission spectra we obtain a single master transmission spectrum $T_{\rm AB}$. The mean $T_{\rm AB}$ of the five observed transits is shown in Fig.~\ref{spettri_trasmissione}. We clearly identify an absorption feature with a FWHM = 0.091$\pm$0.005~nm. The peak of the  excess absorption is measured to be $0.75 \pm 0.03$\,$\%$\,(25$\sigma$), evaluated by fitting a Gaussian (with a local non-linear least-squares method based on the Levenberg–Marquardt algorithm) in the core of the strongest \ion{He}{I} component. The measured absorption levels in individual nights are shown in Table~\ref{table:2_1}. 
	We note that the helium feature has a net blueshift of $-3.0 \pm 0.6$  km s$^{-1}$. This is in agreement with the findings by \citet{Salz2018}, who reported a blueshift of $-3.5 \pm 0.4$  km s$^{-1}$.\\
	
	\begin{table}
		\caption{Average peak absorption in the strongest \ion{He}{I} component in each individual night. }
		\label {table:2_1}    
		\centering        	\begin{tabular}{c c}         
			\hline\hline                        
			Date & \ion{He}{I} absoprtion peak [$\%$]  \\    
			\hline                                  
			transit 1 & 0.64\,$\pm$\,0.10\\
			transit 2 & 0.76\,$\pm$\,0.06\\ 
			transit 3 & 0.96\,$\pm$\,0.07\\
			transit 4 & 0.76\,$\pm$\,0.05\\
			transit 5 & 0.85\,$\pm$\,0.10\\
			\hline              
		\end{tabular}
	\end{table}

	\noindent{Fig.~\ref{tom_he_star} highlights that the helium transmission signal seems to start after $t_1$, and close to $t_2$. This is compatible with an extended atmosphere with a compact structure, i.e. the helium atmosphere is not so elongated as to give a signal already at first contact, but it becomes detectable when the planet is completely inside the stellar disk. This is in agreement with the results of our 3-d simulations (see \S~\ref{EVE}) and with findings by \citet{Salz2018}.}

	\begin{figure*}
		\centering
		\includegraphics[width=17cm]{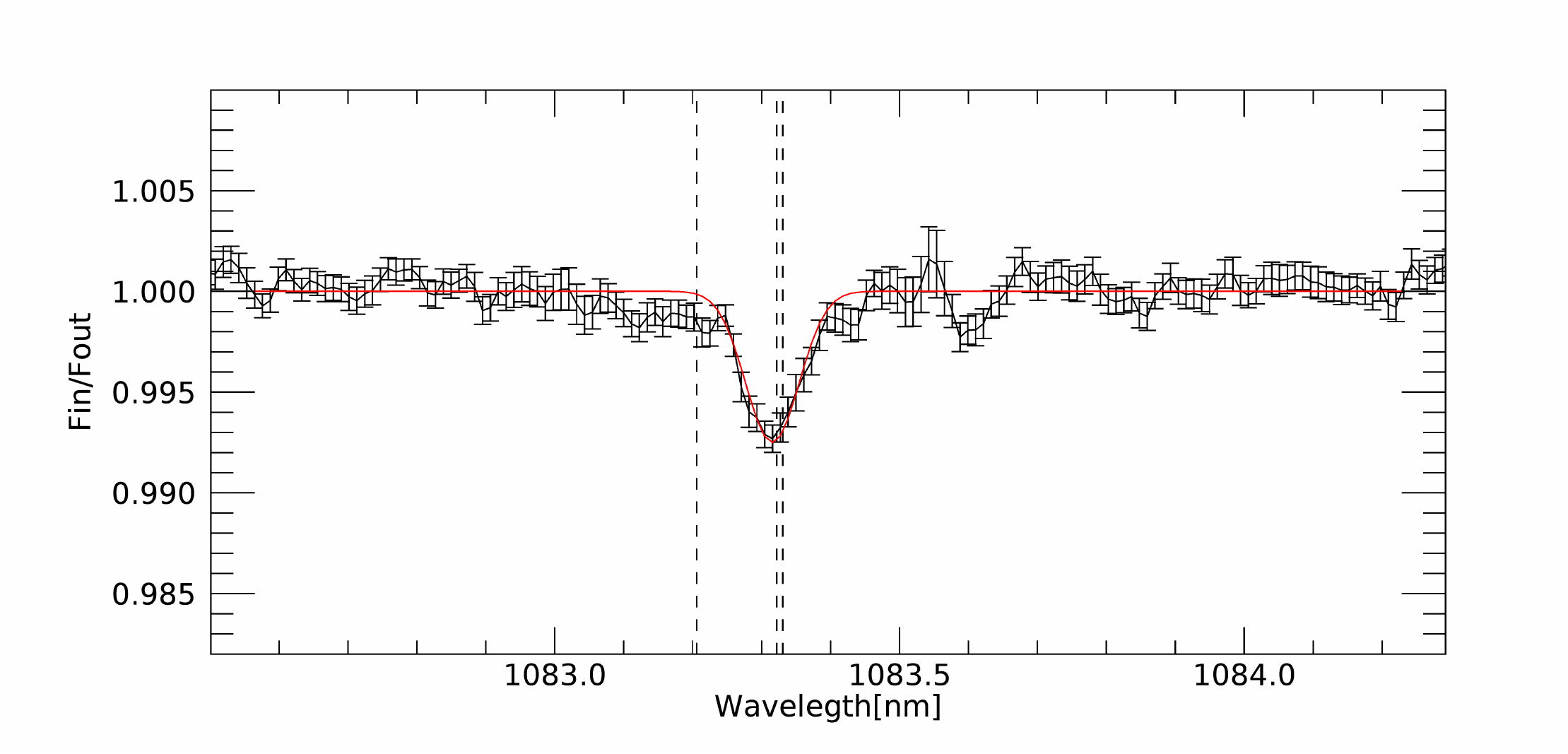}
		\caption{Mean $T_{\rm AB}$ of the five observed nights in the planetary rest frame. We detect an average absorption signal of $0.75 \pm 0.03$\,$\%$\,(25$\sigma$) in the strongest component of the helium triplet. A Gaussian fit is shown as the red line. Vertical dashed lines indicate the three helium triplet lines.}
		\label{spettri_trasmissione}
	\end{figure*}
	
	\subsection{Transmission spectra equivalent widths $\&$ Light-curve analysis} \label{EW_par}
	Once the series of $T_{\rm A}$ and $T_{\rm B}$ have been computed, we evaluate the excess in-transit absorption in the helium triplet lines by using the methodology proposed by \citet{Cauley2015, Cauley2016, Cauley2017, Cauley2017A}, and \citet{Yan2018} for the analysis of the Balmer lines. For every night, the absorption is calculated as the equivalent width of the transmission spectrum at the position of the helium line:
	\begin{equation}
		{EW_{\rm A}}^i=\sum\limits_{v=v_{min}}^{v_{max}} 1-{T_A(v)}^i \;\Delta\lambda_{v}\;, 
	\end{equation} 
	where ${EW_{\rm A}}^i$ is the equivalent width corresponding to the transmission spectrum ${T_{\rm A}}^i$, at the orbital phase $i$, for the nodding position A, and $\Delta\lambda_{v}$ is the wavelength difference at velocity $v$. ${EW_{\rm B}}^i$ is computed similarly for the nodding position B.
	We measure the absorption of the helium triplet across a 40 km s$^{-1}$ band, from $-20 $ km s$^{-1}$ ($\lambda_{min} \sim$ 1083.22~nm) to $+20$ km s$^{-1}$ ($\lambda_{max} \sim$ 1083.36~nm) centred on the strongest component of the triplet. The interval is chosen to avoid possible contamination from the nearby telluric line at 1083.51~nm. 
	
\begin{figure*}
		\resizebox{\hsize}{!}{\includegraphics[width=\linewidth]{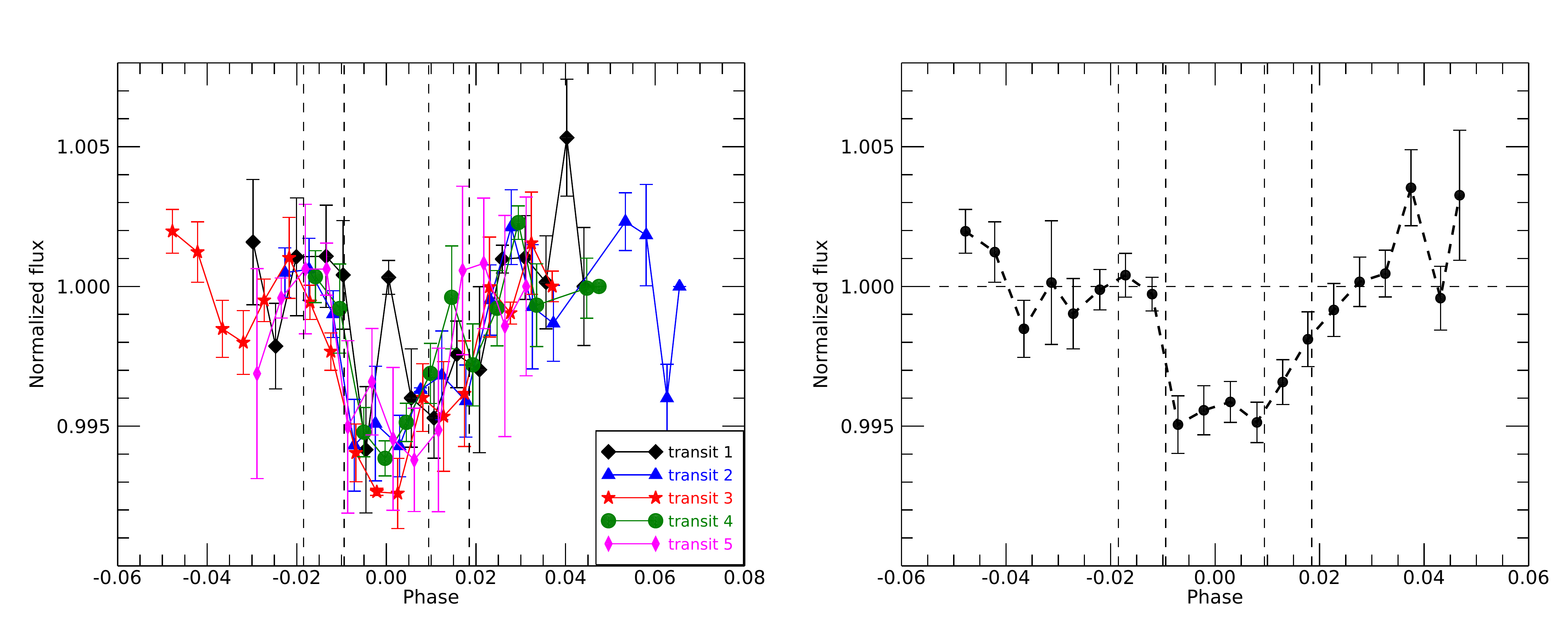}}
		\caption{\textit{Left panel}: Transmission light-curve of the \ion{He}{I} lines in the planetary rest frame from the five transits. The contact point $t_1$, $t_2$, $t_3$ and $t_4$ are marked with vertical dashed lines.\protect\newline \textit{Right panel}: Transmission light-curve, binned in phase, obtained for the five nights combined. The contact points $t_1$, $t_2$, $t_3$, and $t_4$ are marked with dashed lines. The continuum behavior is indicated by the horizontal dashed line.}
		\label{He_spectro}
	\end{figure*}

	All the five nights exhibit an extra absorption during the transit, while the out-of-transit $EW$ values are compatible with zero. The mean in-transit \ion{He}{I} extra absorption due to the extended planetary atmosphere, calculated by merging the two nodding positions and averaging the $EW$ values between the contact points $ t_2 $ and $ t_3 $, is $\overline{EW}= 6.5 \pm 0.6\,\rm{m}\AA$, which corresponds to a 11$\sigma$ detection (the error bar on this value is evaluated as the standard deviation of the mean). 
	\\	
	\\
	Starting from the $T_{\rm A}$ ($T_{\rm B}$) shifted in the planetary rest frame, for each night we subsequently compute a light-curve of the \ion{He}{I} lines as a function of time \citep[e.g.,][]{Charbonneau2002, Snellen2008}. 
	For every orbital phase, we average the corresponding $T_{\rm A}$ ($T_{\rm B}$) over a velocity range from -20 to +20 km s$^{-1}$ centred on the strongest \ion{He}{I} component, analogously to the equivalent width method. 
	In Fig.~\ref{He_spectro} we show the derived light-curves binned in intervals of 0.005 in phase.	Table~\ref{table:3} lists the average nightly \ion{He}{I} transit depths in the light-curves calculated between the contact  points  t$_2$ and  t$_3$.
	\begin{table}
		\caption{Averaged \ion{He}{I} transit depths in each individual night between the contact points t$_2$ and t$_3$ in the transmission light-curves.}
		\label {table:3}    
		\centering                                     
		\begin{tabular}{c c}         
			\hline\hline                        
			Date & Average \ion{He}{I} transit depth [$\%$]  \\    
			\hline                                  
			transit 1 & -0.36\,$\pm$\,0.03 \\
			transit 2 & -0.51\,$\pm$\,0.03 \\ 
			transit 3 & -0.67\,$\pm$\,0.03\\
			transit 4 & -0.50\,$\pm$\,0.03\\
			transit 5 & -0.47\,$\pm$\,0.06 \\
			
			\hline                                            
		\end{tabular}
	\end{table}

	\section{Planetary absorption vs. stellar activity effects}\label{sec:discussion}
	
	Our multi-technique analysis of the GIANO-B transmission spectra of HD\,189733\,b confirms the absorption feature at the wavelength of the \ion{He}{I} nIR triplet. By looking at the tomography map, we have shown that the signal is at rest in the planetary reference frame, and by using different approaches (i.e. transmission spectra equivalent widths technique and light-curve analysis) we have shown that the signal is present in all five transits. The strength of the average absorption signal is measured to be slightly lower (a 2.6$\sigma$ difference) than that ($0.88 \pm 0.04$\%) reported by \citet{Salz2018} in their analysis of CARMENES primary transit spectra of HD\,189733\,b. However, by looking at the left panel of Fig.~\ref{He_spectro} and Table~\ref{table:3}, the night-to-night variability of the \ion{He}{I} absorption levels appears rather clear (a 3.3$\sigma$ and 4$\sigma$ departure for night 1 and 3, respectively, with respect to the quasi-constant values for the other three nights). 
	
	A likely origin for this variability could be the planetary transit on an inhomogeneous stellar surface.  The occultation of active and quiescent regions by the transiting planet can indeed alter the planetary helium absorption signal. The effect on high-resolution transit signatures seen in chromospheric lines is sometimes referred to as the ``contrast effect'', and has been extensively studied in the case of HD~189733b by, for instance, \cite{Barnes2016,Cauley2017}, and \cite{Cauley2018}.  The latter paper, in particular, discusses the contrast effect for several chromospheric lines in addition to the \ion{He}{I} line.
	
	\subsection{The fractional area of active regions on HD~189733} \label{fillingfactor}
	The contrast effect on the transit signal clearly depends on type and extent of non-quiescent (active) regions on the stellar surface. Following \cite{Andretta1995} and assuming that the active regions are similar to solar plage regions, it is possible to obtain a lower limit for the active region area coverage, $f$, through a measurement of the equivalent width of the stellar \ion{He}{I} 1083.3~nm line alone, using the relation:
	\begin{equation}
		f \ge \frac{W_\mathrm{obs}-W_\mathrm{q}}{W_\mathrm{max}-W_\mathrm{q}}\;,
	\end{equation}
	where $W_\mathrm{q}$ is the intrinsic equivalent width in quiescent region, $W_\mathrm{max}$ is the maximum intrinsic equivalent width in active region, and $W_\mathrm{obs}$ is the observed equivalent width of the line\footnote{We adopt here the following definition for the equivalent width measured in the stellar spectra: $W_\lambda = \int (1-(F(\lambda)/F_\mathrm{cont})\:\mathrm{d}\lambda$ where $F(\lambda)$ is the flux line profile and $F_\mathrm{cont}$ is the flux in the adjacent stellar continuum; the equivalent width thus defined is a positive quantity for absorption lines.}.
	We measure the equivalent widths for the off-transit \ion{He}{I} profiles for all the transit dates by means of multi-line fits including the three components of the helium triplet and the main stellar and telluric blends, as discussed by \cite{Andretta2017}.  We obtain values for $W_\mathrm{obs}$ ranging from 320 to 350~m\AA. Adopting the values $W_\mathrm{q} = 40$~m\AA\ and $W_\mathrm{max} = 410$~m\AA\ given by \cite{Andretta2017}, we therefore estimate that the stellar surface of HD\,189733 is covered for at least 75\% by plage-like active regions.  A better estimate of the fractional active region coverage could be obtained by measuring in addition the \ion{He}{I} 587.6~nm line (observed in the HARPS-N range) and computing the \ion{He}{I} 1083.3~nm and 587.6~nm intrinsic equivalent widths from active regions in the case of HD\,189733 over a grid of non-LTE radiative transfer calculations.  Such calculations are outside the scope of this work, although we do plan to include them in a follow-up analysis.
	
	\subsection{Signature of the contrast effect in chromospheric lines} \label{activitysignatures}
\begin{figure*}
		\centering
		\includegraphics[trim=20 40 20 0,angle=90,width=\textwidth]{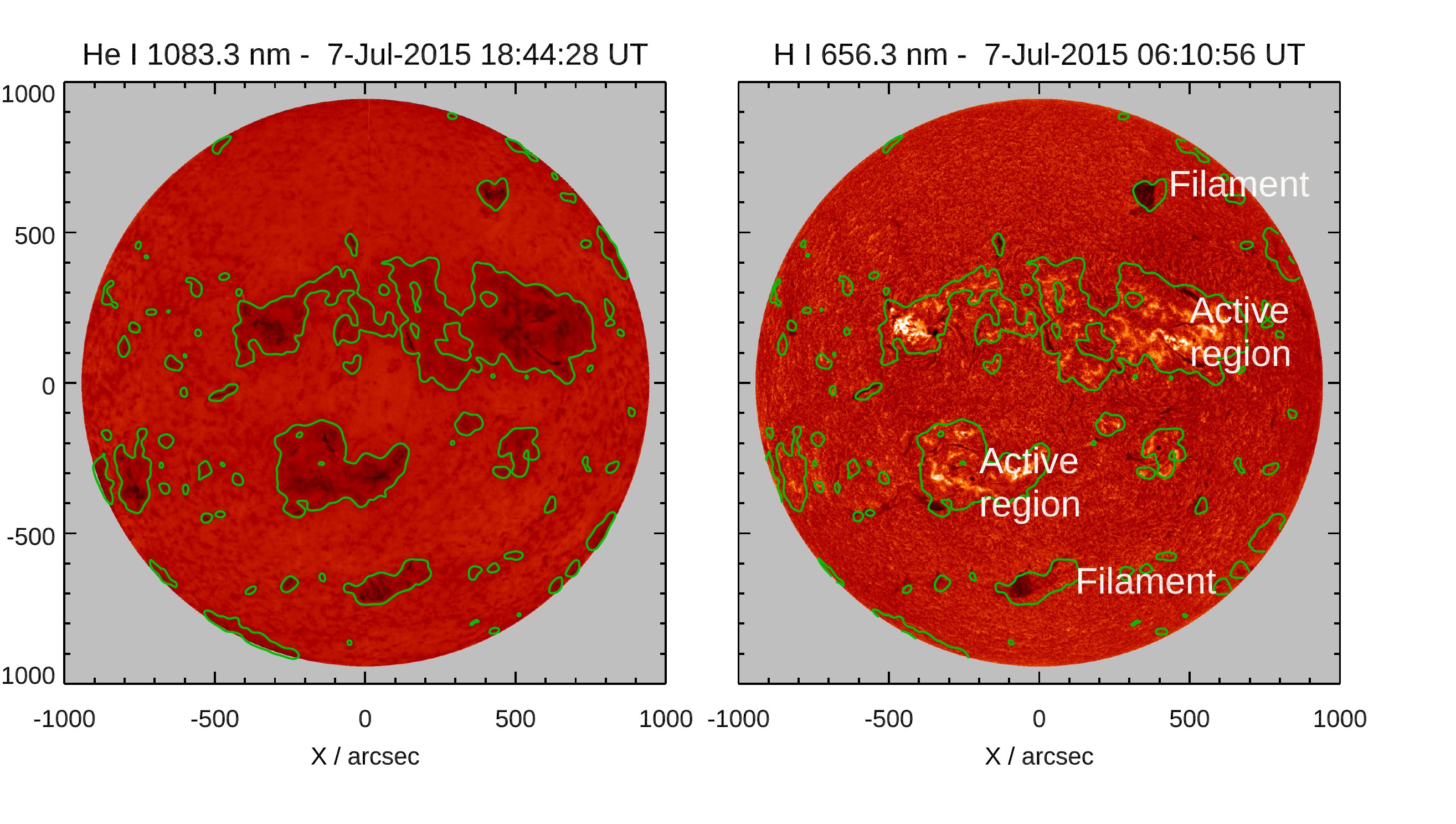}
		\caption{Solar maps corrected for average limb variation of \ion{He}{I} 1083.3~nm (left) and H$\alpha$ (right). Smoothed green contours delineate areas with enhanced emission or absorption with respect to the average line centre intensity.  
		}
		\label{active_region}
	\end{figure*}

	Here we follow instead a different approach, i.e. we look for signatures of the contrast effect in other chromospheric lines in addition to the \ion{He}{I} lines.  One of the most prominent chromospheric line in our data set is the H$\alpha$~656.3~nm line. To illustrate the different effect of stellar surface inhomogeneities on the \ion{He}{I} and the H$\alpha$ lines, we show in Fig.~\ref{active_region} a typical solar image of central line intensity in the two lines.
	The \ion{He}{I}~1083.3~nm map was obtained from the FTP archive of the National Solar Observatory Integrated Synoptic Program (NISP\protect\footnote{\protect\url{https://www.nso.edu/telescopes/nisp} \label{footnote_NSO}}).
	The near simultaneous \ion{H}{I}~656.3~nm map is from Kanzelh\"ohe Observatory and was obtained from the FTP site of the Big Bear Solar Observatory FTP archive of the Global H$\alpha$
	Network\protect\footnote{\protect\url{http://www.bbso.njit.edu/Research/FDHA} \label{footnote_BBSO}}.
	Those maps clearly show that plage-like active regions appear darker than the rest of the stellar disk in \ion{He}{I} and brighter in H$\alpha$. We also note that, on the other hand, filaments appear darker in both lines. %
	This is a typical behaviour of those lines in the Sun and, by extension, in solar-like active stars such as HD~189733.
	
	Considering the \ion{He}{I} line alone, a planet transiting over quiescent areas of the star would increase the weight of active regions in the observed flux, thus producing a stronger 1083.3~nm absorption.  Viceversa, if the planet occults an active region, the net effect is a reduced \ion{He}{I} absorption.  These variations induced by the contrast effect in the in-transit spectrophotometry are very difficult to distinguish from the intrinsic signal due to the planet atmosphere.  \citet{Salz2018} tried to distinguish between these two contributions but did not obtain clear results.
	
	On the other hand, the signal perturbations in the H$\alpha$ line due to the contrast effect have an opposite sign: when the planet occults quiescent regions, the total flux in the line increases, while occultation of active regions produces a decrease of the line flux (assuming that the dominant inhomogeneities are plage-like regions). We also note that the other prominent chromospheric lines in the HARPS-N range, the \ion{Na}{I} D doublet at 589.6~nm (D$_1$) and 589.0~nm (D$_2$) behave very much like the H$\alpha$ line \citep[e.g.,][]{Cauley2018}.
	
	Therefore, if we find evidence that the perturbations in the \ion{He}{I}~1083.3~nm and the H$\alpha$ signals are anti-correlated, we could reasonably infer that the source of the signal is due to the contrast effect.  For this reason, we perform a comparative analysis with the H$\alpha$ line to determine the extent to which our  \ion{He}{I} planetary detection is contaminated by stellar activity, first analysing the out-of-transit line profiles, then the perturbations during the transit of HD~189733b.

	\subsection{Stellar variability among the different nights} \label{stellarvariability}
	In order to estimate the role of stellar activity in the observed \ion{He}{I} transit signal, we analyse the average stellar spectra outside the transit as a function of stellar activity.  In the case of GIANO-B data, we remove the fringing pattern around the \ion{He}{I} line using Method \#2 described in Appendix~\ref{App_A}. We do not attempt to correct the line profiles for the telluric spectrum.
	
	\begin{table}
		\caption{Mean \logRHK\ indexes determined from HARPS-N spectra.}
		\label {table_logRHK}    
		\centering                                     
		\begin{tabular}{c c}         
			\hline\hline                        
			Date & \logRHK  \\    
			\hline                                  
			transit 1 & -4.472 $\pm$ 0.007\\     
			transit 2 & -4.546 $\pm$ 0.005\\
			transit 3 & -4.512 $\pm$ 0.006\\
			transit 4 & -4.500 $\pm$ 0.005\\    
			transit 5 & -4.527 $\pm$ 0.005\\
			\hline                                            
		\end{tabular}
	\end{table}
	
	We first measure the \logRHK\ index from HARPS-N spectra.  The average values for each night are given in Table~\ref{table_logRHK}. From those values, it is clear there is a noticeable variability in stellar activity, with the star being in its higher activity state during transit 1, while transits 3 and 4 exhibit similar but lower levels of activity.  The activity of HD\,189733 is lowest during transits 2 and 5. 
	We also measure the \logRHK\ index for each HARPS-N spectrum, thus obtaining temporal sequences of that index.  Inspection of those light curves does not reveal significant events such as flares. The variations observed during each night are of lower amplitude than the variations of the night-to-night average value given in Table~\ref{table_logRHK}.

\begin{figure*}
		\centering
		\includegraphics[angle=90,width=\linewidth]{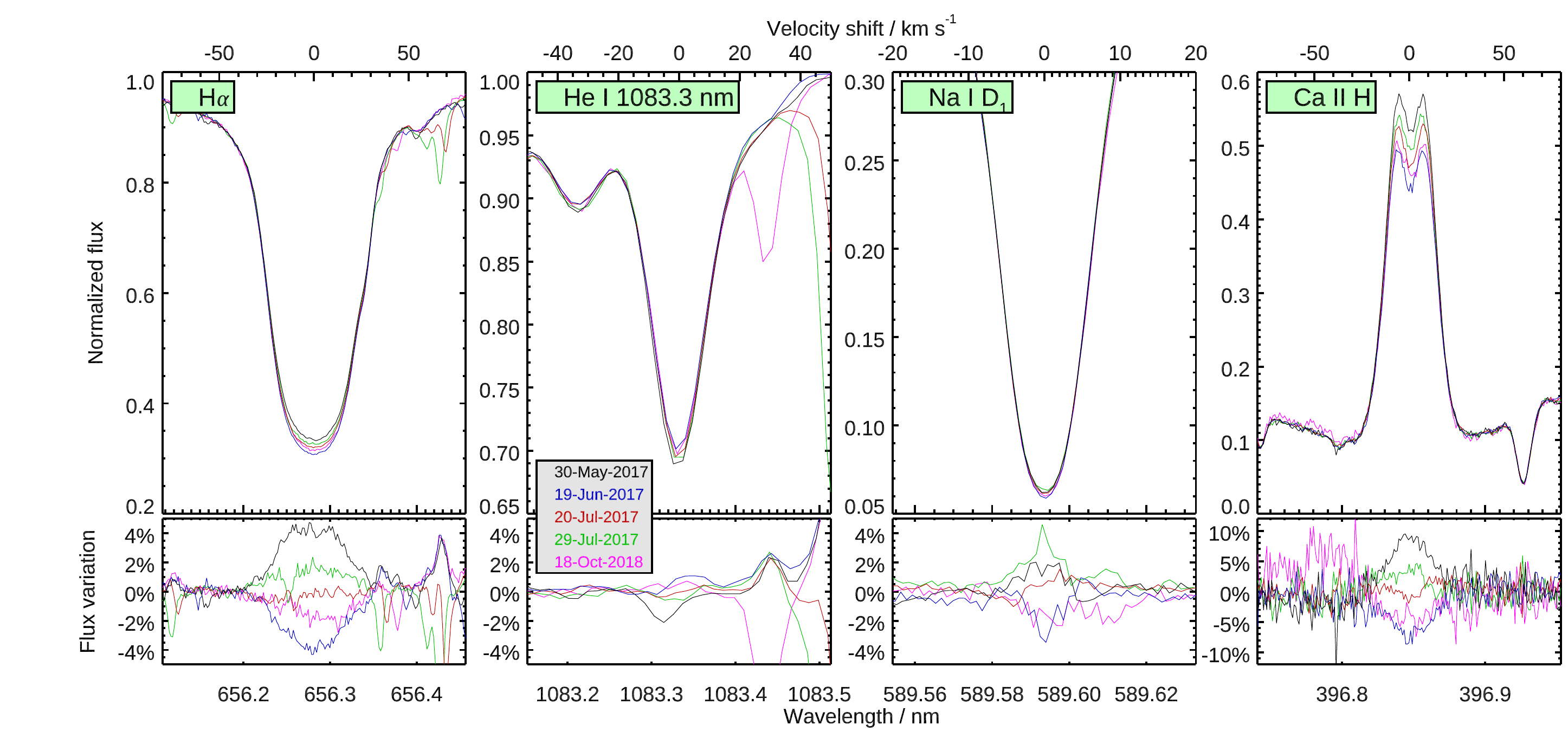}
		\caption{The top panels show, from left to right, the mean out-of-transit profiles of H$\alpha$, \ion{He}{I} 1083.3~nm, \ion{Na}{I} D$_1$, and \ion{Ca}{II}~H.} The mean profile for each date is drawn with the same colour coding as in Fig.~\ref{airmass_}. The bottom panels show the relative variation from the reference, average profile (same colour coding).
		\label{he_ha_v}
	\end{figure*}

	For each transit date, we then compute the average out-of-transit profile of several lines sensitive to stellar chromospheric activity: H$\alpha$, \ion{Na}{I} D$_1$ and D$_2$, \ion{Ca}{II} H and K (all from HARPS-N), and \ion{He}{I} 1083.3~nm (from GIANO-B). Each average line profile is normalized using the wings of the line or the nearby continuum (Fig.~\ref{he_ha_v}, top panels). It should be noted that the integrated \logRHK\ indices listed in Table~\ref{table_logRHK} are mean values for each transit night, while the average profiles shown in Fig.\ref{he_ha_v} are computed from out-of-transit spectra only.
	
	From the set of 5 average out-of-transit profiles, $F_i(\lambda)$ ($i=1\ldots 5$), we compute an average, reference line profile, $F_\mathrm{ref}(\lambda)$.  In the bottom panels of Fig.~\ref{he_ha_v} we show the variation of line fluxes with respect to that reference profile: $R_i(\lambda) = F_i(\lambda)/F_\mathrm{ref}(\lambda)$.  We only show the \ion{Na}{I} D$_1$ line since the D$_2$ line is significantly affected by telluric absorption in some dates. Also, the \ion{Ca}{II} K and H lines behave very similarly, thus we only show the latter.  We notice above all that the relative variations of \ion{He}{I} fluxes are anti-correlated with H$\alpha$ and \ion{Ca}{II}~H: for example when H$\alpha$ displays an excess flux in the line core (as in the profile from transit 1), the helium line shows a deeper core flux.  The \ion{Na}{I} D$_1$ relative flux, although noisier and narrower, follows the trend of the H$\alpha$ line, as it is expected from the analogy with solar observations and in qualitative agreement with the calculations of \citet{Cauley2018}.
	
\begin{figure}
		\includegraphics[angle=90,width=\linewidth]{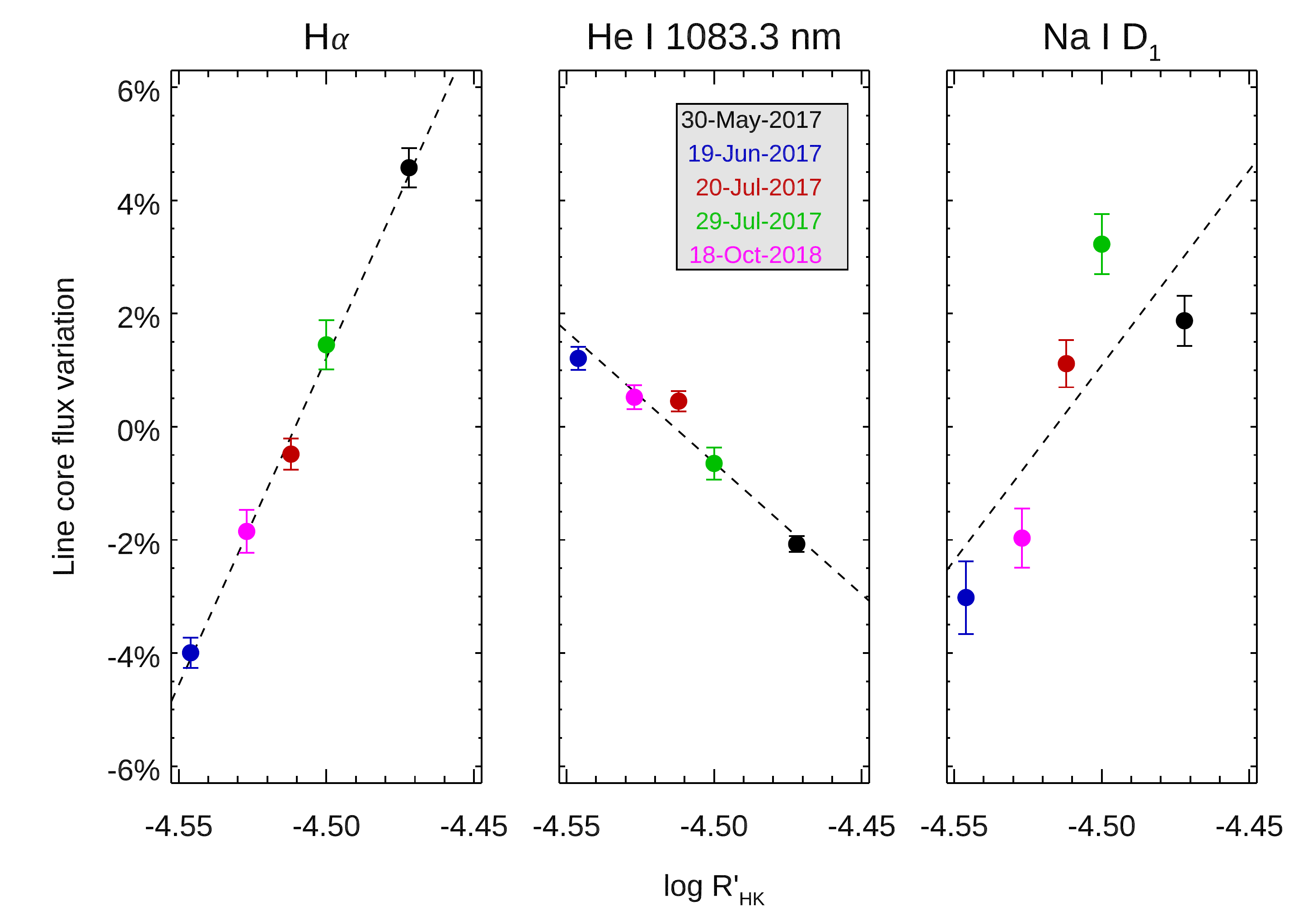}
		\caption{Variation in the central line depth of the mean out-of-transit stellar line profiles as a function of stellar activity as measured by the \logRHK\ index. Linear fits to the measurements are shown with dashed lines.}
		\label{relatio}
	\end{figure}

	To better quantify this effect, we estimate the peak values of $R_i(\lambda)$ in the line core by a Gaussian fit in a range as free as possible from telluric blends.  More specifically, the ranges for the fit procedure are: 656.177 -- 656.345 nm for H$\alpha$, 1083.23 -- 1083.38 nm for \ion{He}{I}~1083.3~nm, and 589.57 -- 589.60 nm for \ion{Na}{I} D$_1$.  The resulting peak values of $R_i(\lambda)$ are shown in Fig.~\ref{relatio} as a function of the activity index \logRHK.
	
	As a sanity check, we also estimate the peak value of $R_i(\lambda)$ in the core of the \ion{Ca}{II}~H line, within the range 396.82 -- 396.87~nm, but we do not display the results in Fig.~\ref{relatio} because they are very well correlated with \logRHK, as expected (correlation coefficient: 0.98).  The correlation coefficients with \logRHK\ for the other lines are tabulated in Table~\ref{table_corrs} along with their $p$-values. We also compute the linear fit of the relative variation of line depth with \logRHK\ (dashed lines in Fig.~\ref{relatio}).  The slopes of the fit is also given in Table~\ref{table_corrs}.  If we interpret these values as the line sensitivity to activity, we can say that, in absolute terms, \ion{He}{I}~1083.3~nm and \ion{Na}{I} D$_1$ have similar sensitivity to \logRHK, while H$\alpha$ is roughly twice as sensitive than both.
	\begin{table}
		\caption{Correlation coefficients of relative line core variation with the \logRHK\ index (in parentheses the corresponding $p$-values), and the slopes of the linear fits.}
		\label{table_corrs}    
		\centering                                     
		\begin{tabular}{lr@{ }lr@{$\pm$}l}         
			\hline\hline                        
			Line & \multicolumn{2}{c}{Correlation coeff.} & \multicolumn{2}{c}{Linear slope} \\    
			\hline                                  
			H$\alpha$                    &  $0.998$ &($0.00011$) &$1.16$ & $0.06$\\     
			\ion{Na}{I}~D$_1$            &  $0.831$ &($0.081$)   &$0.69$ & $0.09$\\
			\ion{He}{I}~1083.3~nm & $-0.970$ &($0.0061$)  &$-0.46$ & $0.03$\\
			\hline                                            
		\end{tabular}
	\end{table}
	
	We notice in particular that there is an excellent, positive correlation between variations in the H$\alpha$ line depth with activity.  A good correlation, but of opposite sign, is seen in the \ion{He}{I}~1083.3~nm too.  Once more, the \ion{Na}{I} D$_1$ variations behave like the H$\alpha$ line, although the data point for the most active night (transit 1) deviates from the linear trend seen in the other 4 nights.
	
	From this analysis, we conclude that the observed night-to-night variations of out-of-transit average profiles are consistent with the picture of a stellar surface covered by large plage-like active regions producing an anti-correlation between \ion{Ca}{II} H \& K (\logRHK), H$\alpha$ (and the \ion{Na}{I} doublet) and the \ion{He}{I}~1083.3~nm line. The effect of stellar activity on HD\,189733's chromospheric lines was also highlighted by \citet{Kohl2018}, who carried out an observing campaign, with the 1.2~m TIGRE (Telescopio Internacional de Guanajuato Robótico Espectroscópico) telescope to monitor
	chromospheric and possibly planet-induced variations in the H$\alpha$ line. They, by studying the time variability of the H$\alpha$ and stellar activity-sensitive
	calcium lines, interpreted the variations measured in the star's H$\alpha$
	profile as dominated by stellar activity.\\
	The effect of an inhomogeneous stellar surface on line profiles during planetary transits will of course depend on the details of the nature and distribution of regions with perturbed line profiles across the stellar disk. The work by \cite{Cauley2018} examines for instances different cases, including a solar-like distribution of active, plage-like regions over ``active latitudes''.  In the latter case, they found that the contrast absorption for the \ion{He}{I}~1083.3~nm line is negative and its absolute value can be as large as 0.2\% (see their Fig.~13).  Under the same assumptions, the contrast absorption for other chromospheric diagnostics are comparable but of opposite sign: for H$\alpha$ is $<0.35$\%, for \ion{Ca}{II}~K is $<0.15$\%, and for \ion{Na}{I}~D$_2$ is $<0.2$\%.  It is therefore entirely possible that the inhomogenous stellar surface of HD~189733 could significantly affect the transmission spectrum during planetary transits, at least for epochs with highest stellar activity.
	
	
	\subsection{In-transit \ion{He}{I} vs H$\alpha$ relation}
	\label{He_Ha_relatio}
	
\begin{figure}
		\centering
		\includegraphics[width=\linewidth]{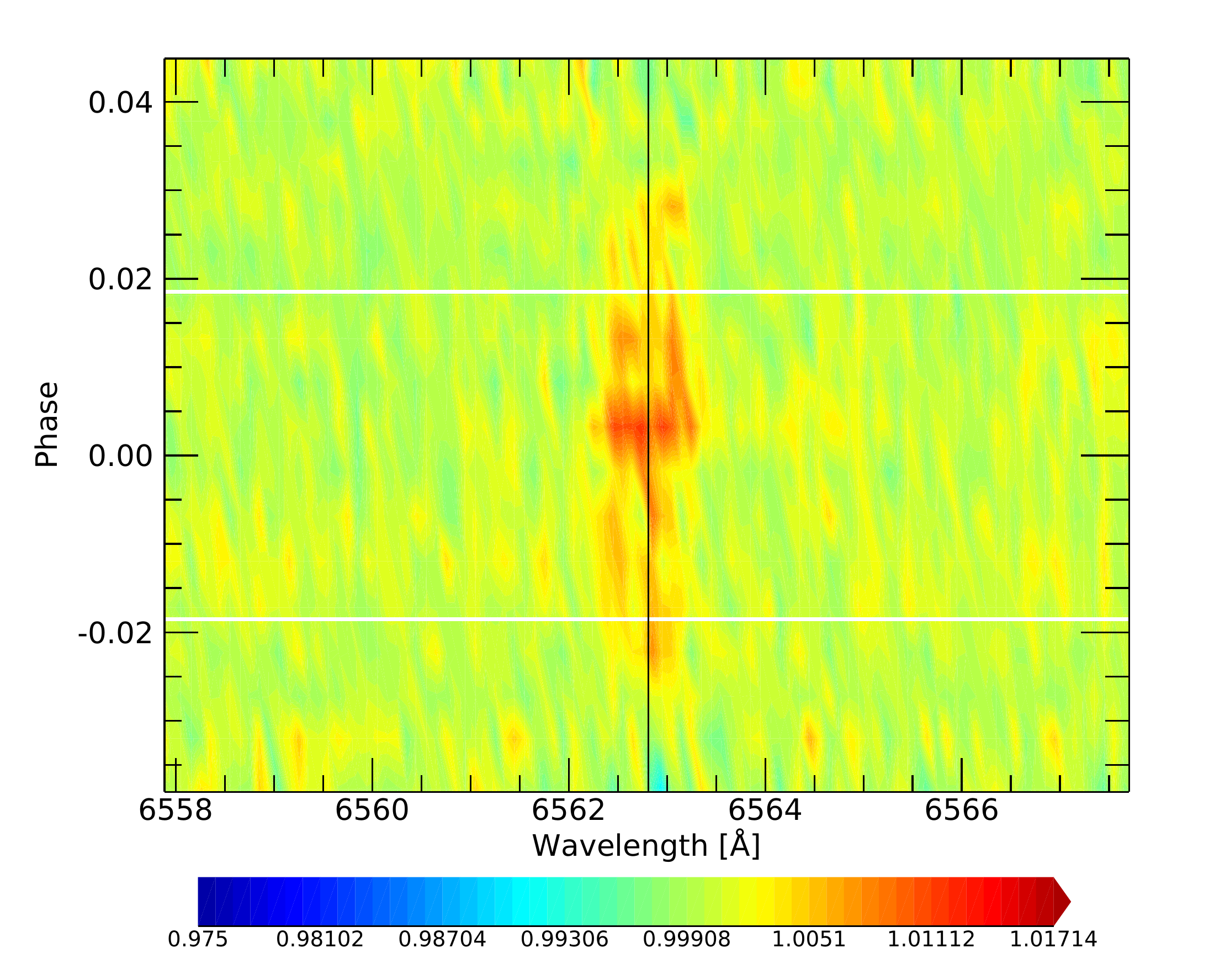}
		\caption{Tomography of the H$\alpha$ line in the stellar rest frame. A strong signal of stellar origin emerges during transit. Horizontal white lines show the beginning and end of the transit.}
		\label{Ha}
	\end{figure}

	Based on the evidence put forth in the previous Section, it appears likely that variability in the planetary helium absorption signal can be due to variations in stellar activity. To put this expectation on more solid grounds, we perform a comparative analysis between the \ion{He}{I} and the H$\alpha$ lines to verify the existence of a correlation. We perform optical transmission spectroscopy on each individual primary transit event using the HARPS-N spectra of HD\,189733\,b in the region around 656.3~nm, following the same method used for the helium nIR triplet (see \S~\ref{Sect3}). By looking at the transmission spectra of all the nights combined together and shown in tomography (see Fig.~\ref{Ha}) a strong emission signal emerges at the position of the H$\alpha$ line, clearly centred in the stellar reference frame.  
	This signal, as it is aligned in the star rest frame, is not due to the planetary atmosphere, but it represents a pseudo-signal caused by the planetary transit over a non-homogeneous stellar disk.
	
\begin{figure}
		\centering
		\includegraphics[width=\linewidth]{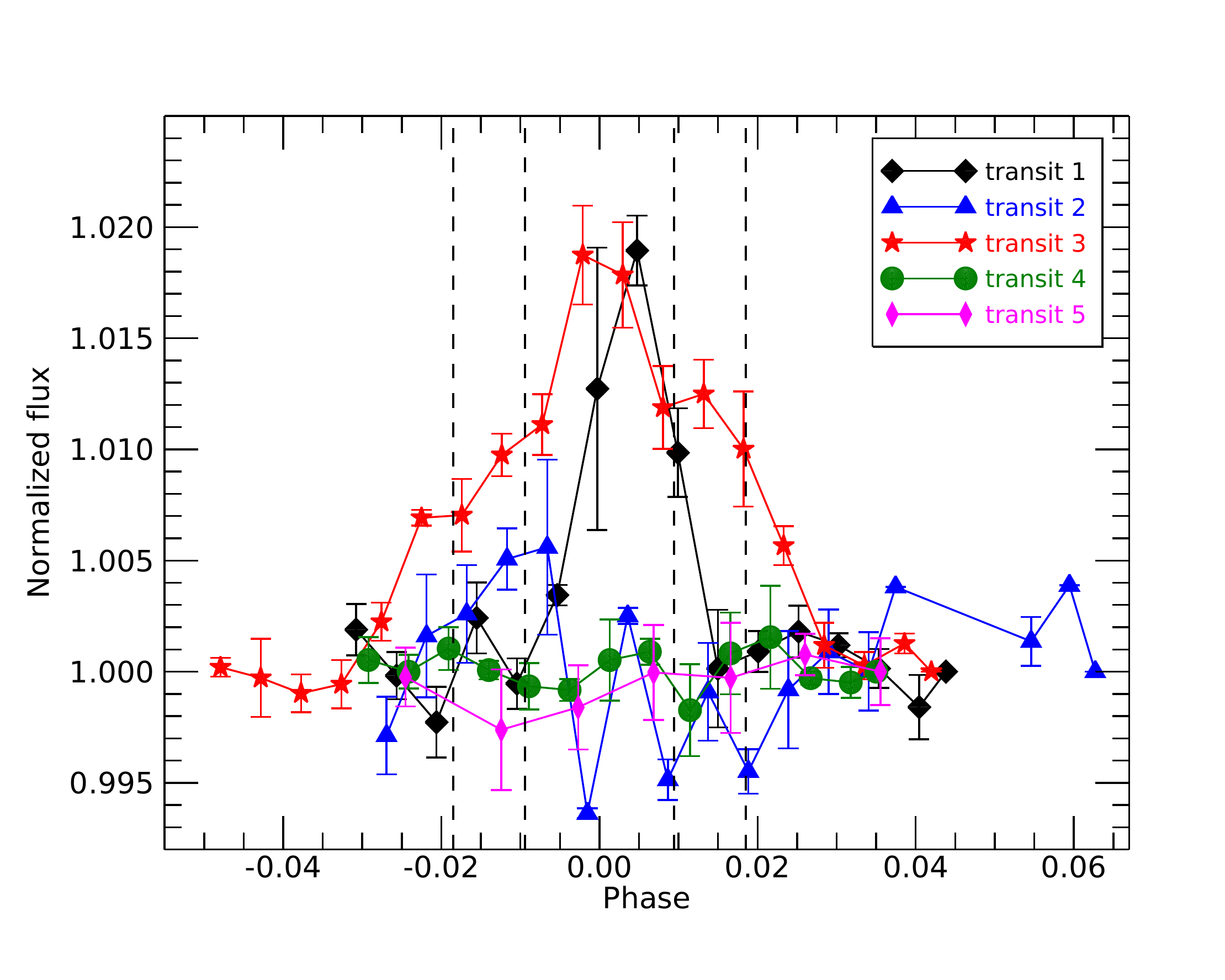}
		\caption{H$\alpha$'s light-curve. Two nights out of five (transit 1 and 3) show the H$\alpha$ signal in emission during transit in the stellar reference frame. }
		\label{Ha_spectro}
	\end{figure}

	By computing the transmission light-curve of the H$\alpha$ line in the planetary rest frame\footnote{We choose the planetary rest frame to be coherent with the helium analysis, but, since we average over a velocity range of $\pm~20 $ km s$^{-1}$, $\pm~0.44~\AA $, centred on the H$\alpha$ line, the results do not change if we derive the light-curves in the stellar rest frame.}, we find that the H$\alpha$ emission signal shown in Fig.~\ref{Ha} comes from two nights out of five (transit 1 and 3), while the other three exhibit constant H$\alpha$ levels in- and out-of-transit (see Fig.~\ref{Ha_spectro}).  
	
\begin{figure}
		\includegraphics[width=9cm]{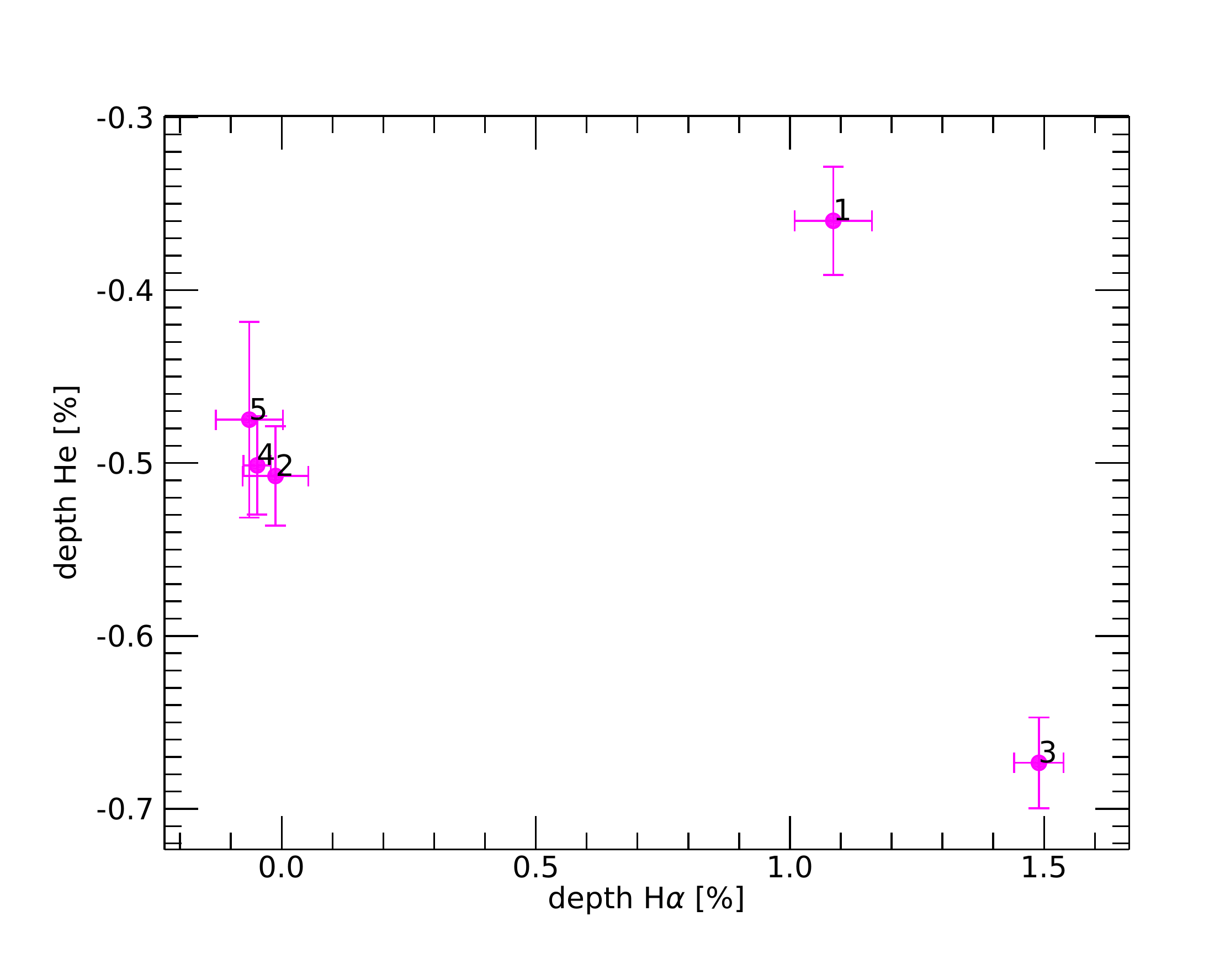}
		\caption{He vs. H$\alpha$ transit depths. When no in-transit variability of the H$\alpha$ signal is detected, the He transit depth is constant.}
		\label{He_ha}
	\end{figure}

	Next, we evaluate night averages of the H$\alpha$ light-curves between contact points t$_2$ and t$_3$. If \ion{He}{I} absorption levels are indeed linked to variability in  H$\alpha$ emission, we would qualitatively expect, for example, stronger absorption levels recorded during transits 1 and 3, in anti-correlation with the stronger H$\alpha$ emission measured. According to the scenario outlined in \S~\ref{activitysignatures}, this would indicate that the planet was transiting over quiet regions of the stellar surface. As shown by Fig.~\ref{He_ha}, this is exactly what we see in the third night, meaning that during the transit the planet could have occulted quiescent stellar regions. Moreover, from Fig.~\ref{Ha_spectro} we note that the H$\alpha$ emission signal measured during this transit event seems to last longer than the actual transit. However, we prefer not to draw too speculative conclusions, as we cannot exclude it is simply due to stellar activity by chance aligned with transit.\\ 
	A different behaviour is instead the one seen during transit 1, as Fig.~\ref{He_spectro} shows. During transit 1 we do not measure increased absorption at the position of the helium triplet -in correspondence of an emission in H$\alpha$-, but a peak in emission in the middle of the transit\footnote{As a consistency check, a very similar behaviour is seen by using the Sodium doublet as a proxy instead of H$\alpha$. See Appendix~\ref{App_B}.}.  
	In principle, an emission signal at the position of both lines could be seen if the planet during the transit is passing over a region that is darker than the rest of the stellar disk, both in \ion{He}{I} and in H$\alpha$. Such an active region could be either a starspot or a filament. To test this possibility, we first check the RME curve of the first night, to see if distortions could be present in the RME profile that could be induced by starspot-crossing, but to no avail. To further explore the possible nature of the active region which may have been hidden by the planet during transit 1, we consider the strong \ion{Si}{I} 1083.0~nm line. We choose a photospheric rather than a chromospheric line because the latter would also be influenced by the presence of a filament -not only by a starspot as a photospheric line (e.g. the \ion{Si}{I} line)- complicating any interpretation on the nature of the occulted region. We would expect the \ion{Si}{I} 1083.0~nm line to grow weaker in case of starspot-crossing events, while the line strength should be unaffected by the planet hiding a filamentary active region.  An analysis of the \ion{Si}{I} 1083.0~nm line (not shown) does not highlight any changes in its strength during transit 1. We therefore tentatively conclude that the active region occulted during the transit could be a filament.

	\medskip
	We need to stress that in this work we do not take into account the possible planetary contribution to the recorded emission levels for the H$\alpha$ line. Indeed, our aim is to understand if a qualitative relation could exist between the \ion{He}{I} and the H$\alpha$ lines, and not to perform a fit between the emission/absorption values we find. Nevertheless, the analysis presented here allows us to infer the fact that three out of five transits of HD\,189733\,b were mostly unaffected by stellar activity, therefore the measured value of planetary \ion{He}{I} absorption could be representative of the real one. We make use of this assumption in the next section.

	\section{Interpretation of the helium absorption and constraints on the mass loss rate}\label{EVE}
	
	We select the HD\,189733\,b transmission spectra obtained for the three transit events not significantly affected by stellar activity
	(transit 2, 4, and 5). The average in-transit absorption depth for the three events is 0.77$\pm$0.04 $\%$ (19$\sigma$), a value in excellent agreement (at the $\sim$0.3$\sigma$ level) with the one calculated considering all the transit events of our dataset (0.75$\pm$0.03). We  interpret our findings in terms of planetary helium absorption using the EVaporating Exoplanet code  \citep[EVE;][]{Bourrier2013b,Bourrier2016}, to take full advantage of the temporal and spectral resolution of the data, and to account for the impact of the stellar surface properties. We use an upgraded version of the EVE code previously applied to observations of planetary helium lines in \citet{Spake2018,Allart2018} and \citet{Allart2019} (we refer the reader to these publications for a complete description of the modelling). 
	
	The  planetary  system  is  simulated  in 3-d  in  the  stellar  rest  frame,  and  the  code  calculates  theoretical spectra comparable to the GIANO-B observations during the transit of the planet. The model star is tiled with the master-out GIANO-B spectrum, scaled in flux according to the continuum limb-darkening in the helium range, and Doppler-shifted according to the orbital motion of the system and the rotational velocity of the star (an obliquity of $\lambda$ = -0.42 deg, v$_\mathrm{eq}$ = 4.45\,km\,s$^{-1}$, i$_{*}$ = $\ang{92.5}$, \citealt{Cegla2016}). We note that in using the master-out as a proxy for the local stellar lines, we assume that centre-to-limb variations (CLV) can be neglected along the transit chord. This choice is motivated by the absence of spurious residuals induced by CLV in the transmission spectra calculated with the master-out (Fig.~\ref{tom_he_star}), and our preference to use the helium spectrum measured directly for HD\,189733 rather than theoretical local stellar spectra that we cannot compare with observations. The use of the EVE code allows us to account for the effect of the RME, for limb-darkening, and for the partial occultation of the stellar disk during ingress/egress. 
	Simulations can be compared directly with the observed spectra, with no need to correct them for these effects.
	\begin{figure}
		\centering
		\includegraphics[width=\linewidth]{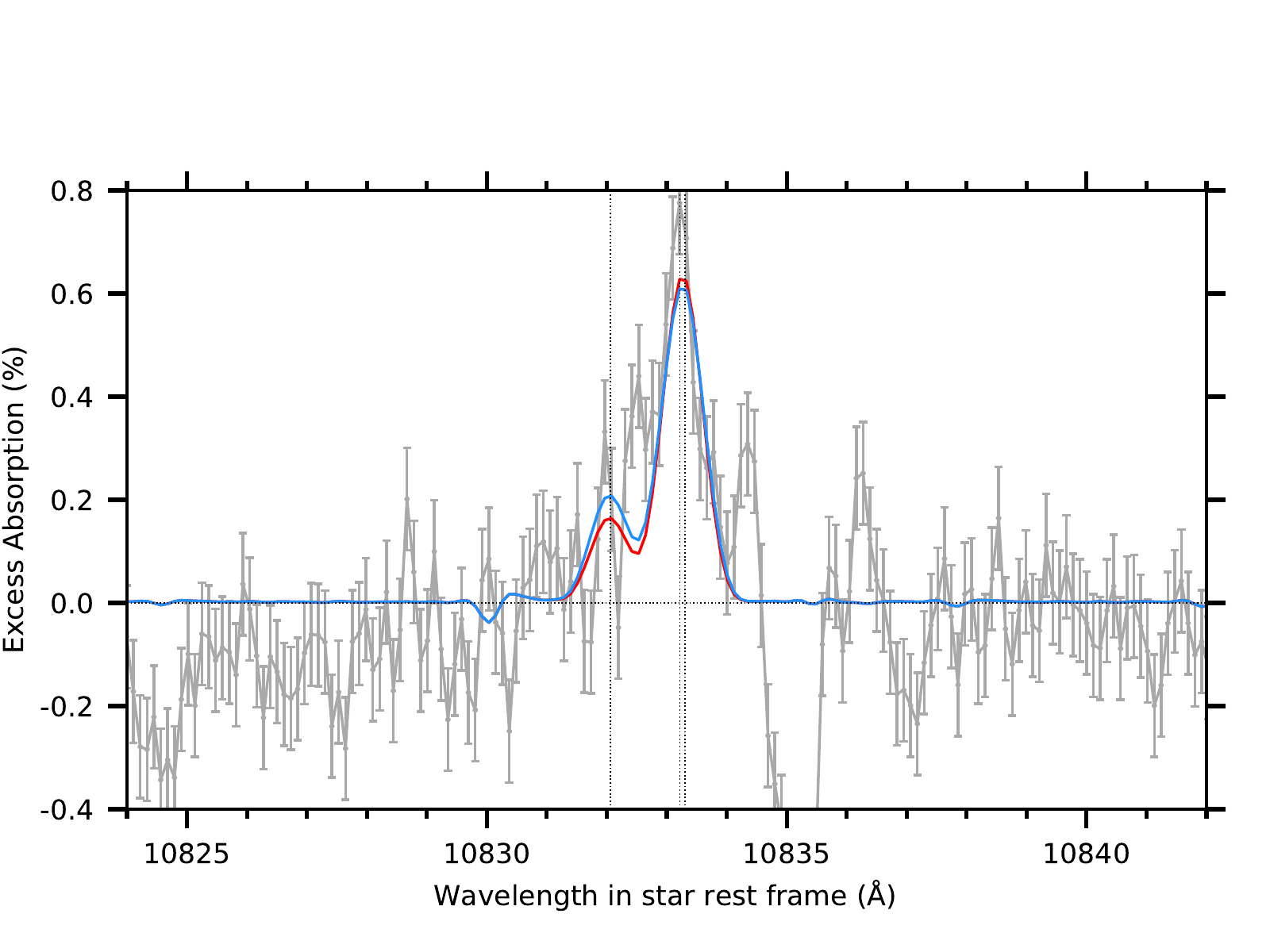}
		\caption{Average excess absorption profile (in the planetary rest frame) from the exposures fully in-transit. Blue profile is the best fit for 1.2 R$_{\rm{P}}$, while the red profile is the best-fit for 3 R$_{\rm{P}}$. Only the three nights that show no stellar contamination (transits 2, 4, and 5) are taken into account.}
		\label{1D_spec_EVE}
	\end{figure}
	
	The thermosphere is  modelled  as  a  parameterized  grid,  using  density  and  velocity profiles calculated with a spherically symmetric, steady-state isothermal wind model \citep{Parker1958}. We assume a solar-like hydrogen-helium composition for the thermosphere and set its mean atomic weight to 1.3. In this first interpretation of the GIANO-B spectra we scale the density profile of metastable helium to the density profile of the wind. In future simulations we will calculate more precisely the populations of ground-state and metastable helium \citep{Oklop2018}. With these settings, the free parameters in the model are the thermosphere temperature, the mass loss of metastable helium, and the extension of the thermosphere. Compared to previous simulations we account for the rotation of the planet by conferring an angular velocity of 3.28$\times$10$^{-5}$s$^{-1}$ (set by assuming tidal-locking) around the normal to the orbital plane to gas in the thermospheric grid.
	
	The theoretical and observed transmission spectra are compared over the spectral range 1082.7 - 1083.7~nm (in the star rest frame), simultaneously fitting the three nights that show no stellar contamination (transits 2, 4, and 5). We calculate grids of temperature versus mass loss, for an extension of the thermosphere of either 1.2 or 3 R$_{\rm P}$ (i.e., the height of the thermospheric grid is either 0.2 or 2R$_{\rm P}$). This choice is motivated by the interpretation of CARMENES observations of HD\,189733\,b by \citealt{Salz2018}, which suggest that the helium lines are saturated and could arise from a compact thermosphere well within the Roche lobe at 3 R$_{\rm P}$.
	Fig.~\ref{1D_spec_EVE} shows the average excess absorption profile (in the planetary rest frame) from the exposures fully in-transit. Blue profile is the best fit for 1.2 R$_{\rm{P}}$, while the red profile is the best-fit for 3 R$_{\rm{P}}$.
	The best-fit (performed by masking the telluric line at 1083.5~nm) yields a reduced $\chi^2$ of about 1.18, it is obtained for an extension of 1.2 R$_{\rm{P}}$, a mass loss of metastable helium of $\sim1-2$~g\,s$^{-1}$ (1$\sigma$), and a temperature of 12000 K.  
	
\begin{figure}
		\centering
		\includegraphics[width=\linewidth]{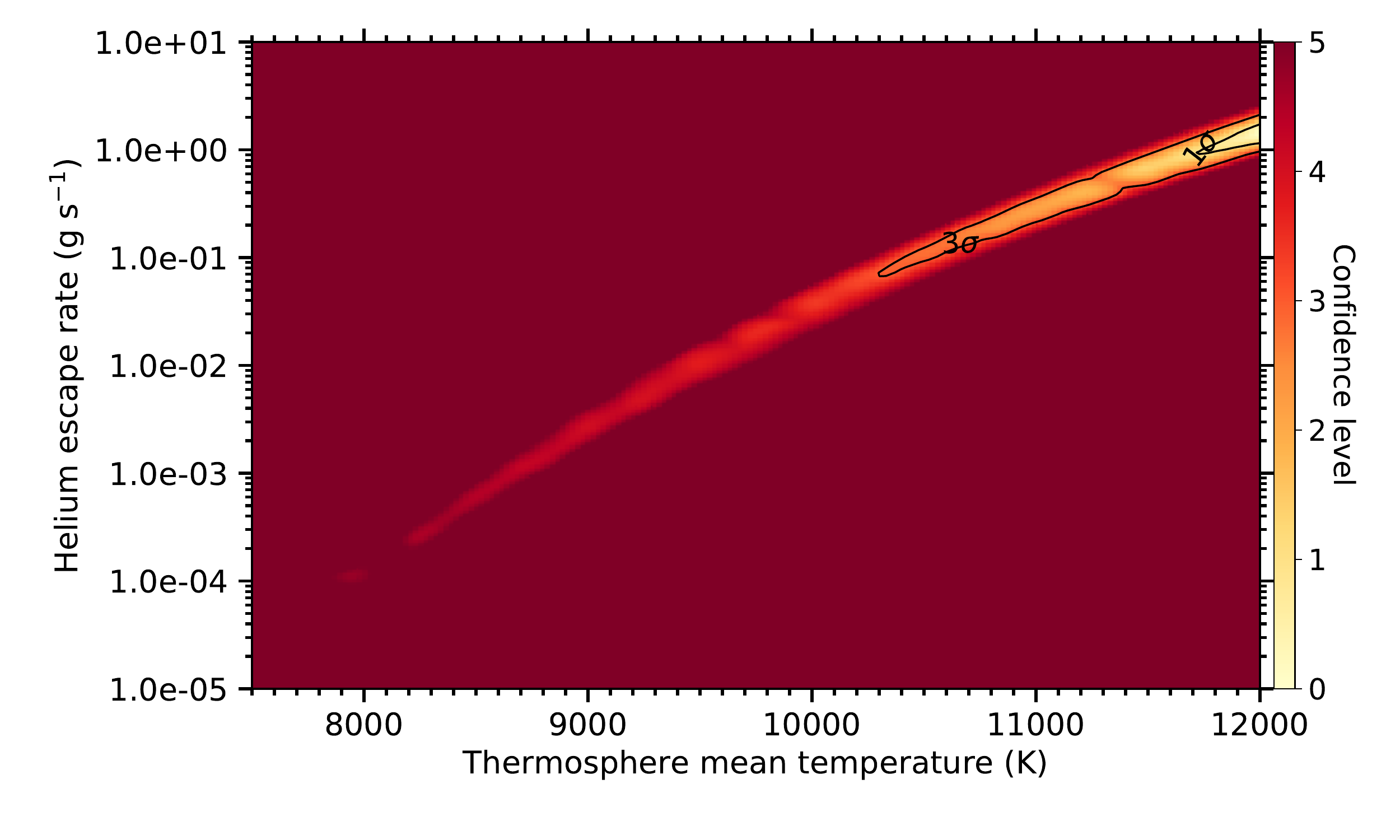}
		\caption{$\Delta\chi^2$ grid (in confidence level) for the 1.2 R$_{\rm P}$ thermosphere extension. Black solid lines indicate the 1 and 3 $\sigma$ confidence contours. The best-fit is obtained for a mass loss of metastable helium of $\sim$ 1-2 g s$^{-1}$ and a temperature of 12000\,K.}
		\label{model}
	\end{figure}

	The data favors hotter thermospheres, but we do not consider temperatures larger than 12000 K because it is at the upper limit expected for HD\,189733\,b \citep[see][]{Salz2016}. Fig.~\ref{model} shows the $\Delta\chi^2$ grid (in confidence level) for the 1.2 R$_{\rm{P}}$ extension. The best fit for the 3~R$_{\rm{P}}$ extension is disfavored with 3$\sigma$ confidence, in agreement with the \citealt{Salz2018} results. 
	
	We confirm that the RME has a negligible impact on transmission spectra of HD\,189733\,b in the region of the helium lines. Spurious signatures induced by the RME are visible in the theoretical spectrum at the location of deep stellar lines, with the W-like shape expected for an aligned system, but their amplitude remains well within the photon noise (Fig.~\ref{1D_spec_EVE}). 
	We also note that tidally-locked rotation has a negligible contribution to the broadening of the helium lines. Our best-fit model has a density of metastable helium of 70 atoms per cm$^{3}$ at 1.2 R$_{\rm{P}}$. This is comparable to the density levels simulated between 1 and 10 R$_{\rm{P}}$ by \citep{Oklop2018} for the hot Jupiter HD\,209458\,b.

	As shown by the tomography map in Fig.~\ref{tom_he_star}, and the light-curve measurements in Fig.~\ref{He_spectro}, the helium absorption signal seems to start immediately after the transit ingress (t$_2$ point) and last after the egress (t$_4$ point). This suggests that the planet could have an asymmetrical cloud of helium, with a small tail (R$_{\rm{tail}}$ <R$_{\rm {Roche}}$) that follows the planet, as found for WASP-69b \citep{Nortmann2018}.
	Moreover, an isotropic, spherical thermosphere model like the one we used, highlights the possible presence of an excess absorption in the blue wing of the lines. We should be cautious about this (because there is no indication of this feature in the CARMENES data), but if confirmed, it would suggest (along with the temporal asymmetry we mentionned) the presence of escaping helium atoms blown away from the planet, a scenario that could become the objective of future work.

	\section{Summary and Conclusions}\label{sec:conclusions}
	\par
	We analyse five transits of HD\,189733\,b acquired with HARPS-N and GIANO-B simultaneously operated in GIARPS observing mode at the TNG. We employ a multi-technique approach (i.e. transmission spectroscopy and tomography, transmission spectra equivalent widths, light-curve analysis) to confirm the recent detection of helium in the planet's extended atmosphere by \citet{Salz2018}. We spectrally resolve the \ion{He}{I} triplet at 1083.3~nm, and detected an absorption signal in each individual night. The in-transit absorption depth is measured to be $0.75 \pm 0.03$\,$\%$\,(25$\sigma$) in the strongest component of the helium triplet. The strength of the average absorption signal is slightly lower that the one ($0.88 \pm 0.04$\,$\%$) reported by \citet{Salz2018}. A possible origin of this difference could be due to the lower resolving power of GIANO-B (R$\sim$50~000) compared to that of CARMENES (R$\sim$80~400). This could in principle cause extra broadening of the lines and thus a decrease in the lines' depth. However, we are not in a position to estimate how much this effect amounts to. Investigating in detail the consequences of having two different resolving powers would require a dedicated simulation, which goes beyond the scope of the paper. Moreover in our analysis a night-to-night variability of the helium absorption levels appears rather clear (at the level of 3.3$\sigma$ and 4$\sigma$ for night 1 and 3, respectively).
	We interpret this variability as likely due to the planet transiting inhomogeneities of the active stellar surface. The occultation of quiescent and active regions by the planet during transit can indeed modify the planetary helium absorption signal. This effect is sometimes
	referred to as the “contrast effect'', and we look for its signatures in other chromospheric lines in addition to the \ion{He}{I} triplet, that is the H$\alpha$, the \ion{Ca}{II} H and K, and \ion{Na}{I} doublet. If the star presents active regions and the planet transits without hiding them, we expect the weight of active regions in the observed flux to increase, producing a stronger absorption at 1083.3~nm. On the contrary, when the planet passes over active regions,
	an emission feature is produced in the in-transit transmission
	spectrum at the position of the \ion{He}{I} lines, reducing the helium absorption. On the other hand, the signal perturbations in the H$\alpha$ (in the \ion{Na}{I} doublet and in the \ion{Ca}{II} H and K lines) have an opposite behaviour. When the planet hides active regions the total flux in line decreases, while occultation of quiescent regions leads to line flux increase (assuming that the dominant inhomogeneities are plage-like regions). \\
	In order to estimate the role of stellar activity in our \ion{He}{I} signal, we first analyse the average stellar spectra outside the transit as a function of stellar activity. From this analysis we find that a nightly variability exists between these out-of-transit spectra and it is consistent with a stellar surface covered by large plage-like active regions.\\
	Encouraged by this empirical evidence, we develop a new approach to evaluate the effects of pseudo-signals induced by stellar activity on the true planetary absorption. The new methodology is based on a comparative analysis of the \ion{He}{I}~1083.3nm (in the infrared) and the  H$\alpha$~(in the visible) lines.
	The results of this analysis are the following:
	\renewcommand{\labelitemi}{$\bullet$}	
	\begin{itemize}
		\item constant helium absorption signals are recorded in three nights out of five (transit 2, 4, and 5), in correspondence of constant H$\alpha$ levels in- and out-of-transit;
		\item one night (transit 3) shows extra absorption in the \ion{He}{I} triplet in connection to an emission signal in the H$\alpha$ line;
		\item one transit (transit 1) exhibits a peak in emission during mid-transit both in helium and in H$\alpha$.
	\end{itemize}
	
	We interpret the three measurements of constant helium absorption levels (0.77 $\pm$ 0.04 $\%$ on average) as mostly unaffected by stellar activity-induced effects, and therefore representative of the actual absorption by the planetary atmosphere. 
	The anti-correlation observed between the \ion{He}{I}~1083.3~nm and H$\alpha$ signals during transit 3 can be explained as due to the contrast effect of the presence of active regions (plages) on the stellar disc not covered by the planet during transit. The peaks in emission recorded in transit 1, both at the position of the \ion{He}{I} triplet and H$\alpha$, can instead be explained as due to the planetary occultation of a stellar filament during transit.
	
	We explain the night-to-night variations in the \ion{He}{I} absorption signal as a consequence of the stellar surface geometry -the presence of quiescent and active regions during planetary transit- and not due to stellar variations, i.e. interaction with stellar wind or flares, change in the X-ray and EUV irradiation. Viceversa, variation in the stellar wind and XUV spectrum could instead influence the density of neutral hydrogen resulting in a temporal variation in the in-transit Ly-$\alpha$ signal \citep{Lecavelier2012,Bourrier2020}. Our feelings are corroborated by the detection of H$\alpha$ in emission -during transit 1 and 3- and not in absorption, as expected by interaction with stellar wind or by variation in the XUV spectrum.
	
	We interpret the HD\,189733\,b helium absorption observations not significantly affected by stellar contamination (transit 2, 4, and 5) in terms of mass-loss rate using 3-d numerical simulations with the EVE code. Our observations are well explained by a compact thermosphere heated to $\sim$12000~K, extending up to 1.2 planetary radii, well within the Roche lobe \citep[in agreement with previous results by][] {Salz2018}, and losing about $\sim1-2$~g\,s$^{-1}$ of metastable helium.
	\medskip
	
	In conclusion, active host stars of hot planets such as HD\,189733 are promising candidates to search for planetary \ion{He}{I}~1083.3~nm absorption from the escaping/extended planetary atmosphere, but absorption in the nIR helium triplet lines due to stellar activity features is a source of serious complications in the interpretation of exoplanet transit observations for active stars. It is therefore of paramount importance to systematically investigate the \ion{He}{I} triplet lines alongside diagnostics of stellar activity in other wavelength regimes (e.g., H$\alpha$ in the optical), using approaches such as the one developed for this work. Only in this way interpretative degeneracies between actual atmospheric absorption and stellar pseudo-signals can be removed and well-determined mass-loss rates obtained from the measured \ion{He}{I} absorption. Hence, the results reported here stress the necessity of simultaneous optical+nIR monitoring when performing high-resolution transmission spectroscopy of hot planets' extended and escaping atmospheres using facilities such as GIARPS.

	\begin{acknowledgements} 
		We want to thank the anonymous referee for the constructive comments which helped to improve the quality of the manuscript.
		We acknowledge the support by INAF/Frontiera through the “Progetti Premiali” funding scheme of the Italian Ministry of Education, University, and Research. 
		G.G. acknowledges the financial support of the 2017 PhD fellowship programme of INAF.  P.G. gratefully acknowledges support from the Italian Space Agency (ASI) under contract 2018-24-HH.0.
		F.B. acknowledges financial support from INAF through the ASI-INAF contract 2015-019-R0.
		V.B. acknowledges support by the Swiss National Science Foundation (SNSF) in the frame of the National Centre for Competence in Research ``PlanetS''. 
		This project has received funding from the European Research Council (ERC) under the European Union's Horizon 2020 research and innovation programme (project Four Aces, grant agreement no. 724427; project Exo-Atmos grant agreement no. 679633).
		M.E.   acknowledges   the   support   of   the   DFG   priority   program
		SPP  1992  "Exploring  the  Diversity  of  Extrasolar  Planets"  (HA 
		3279/12-1).
	\end{acknowledgements}  
	
	\bibliographystyle{aa}
	\bibliography{refelio}
	
	\begin{appendix}
		\section{The fringing correction} \label{App_A}
		Fringes in nIR observations often arise from the interference of the incident and internally reflected beams within the thin layers of the detector.  In particular, the GIANO-B detector (a Rockwell Scientific HAWAII-2 2048$\times$2048 imaging sensor) includes a 0.38~mm sapphire substrate which may generate a near-sinusoidal pattern with frequency of approximately 7.5~cm$^{-1}$ in the nIR (see note \ref{footnote_1}). There is currently no standard procedure to account for this effect in the standard GIANO-B data processing pipeline. We therefore explore two different methods to remove this pattern from the spectra in the vicinity of the \ion{He}{I}~1083.3~nm line, one focused on correcting this effect at the level of the transmission spectra and the other applied on the original spectra.
		
		\subsubsection*{Method \#1}
		We correct our $T_{\rm A}$ ($T_{\rm B}$), in the stellar rest frame, as follows:
		\begin{itemize}
			\item For every night, and for each nodding position, we divide the $T_{\rm A}$ ($T_{\rm B}$), in the stellar rest frame, into 5 groups:
			\begin{enumerate}
				\item before-transit;
				\item ingress (between the contact points $t_1$ and $ t_2 $);
				\item in-transit (betwenn the contact points $t_2$ and $ t_3$);
				\item egress (between the contact points $t_3$ and $ t_4$);
				\item out-of-transit.
			\end{enumerate}
			\item For each of these groups, we build a reference spectrum by averaging on time, and we fit the fringing pattern using the following function: $F_{frin}= A[0] * \sin{A[1]*\lambda+a[2]} + A[3]+a[4]*\lambda$. 
			Once for each group the mean fringing pattern fit ($F_{frin}$) has been calculated, we divide each spectrum of each group for the corresponding $F_{frin}$.
		\end{itemize}
		Fig.~\ref{fringing} shows, for transit 1 and 3, the mean in-transit transmission spectra for the A-nodding position before the fringing correction (left panels) with overplotted $F_{frin}$ (in red), and after the fringing removal (right panels).
		
		We wish to underline that the choice to split the transmission spectra into 5 groups is made a priori to account for any possible time-dependent intra-night fringing variations that might have been introduced during the initial steps of the analysis. In case of a constant fringing pattern throughout the night, the net result is a slight increase in the noise on the fringing correction, with no direct impact on the results of our analysis.

		\subsubsection*{Method \#1b}
		
		As a consistency check of Method \#1, we also calculate, for each night, the Fast Fourier Transform (FFT) in the wavelength domain of each transmission spectrum $T_{\rm A}$ ($T_{\rm B}$). We then select the two most prominent frequencies of the Fourier spectrum and for them we define the module of the FFT equal to zero.
		
		We then perform an inverse transform to bring the data back into the wavelength domain, filtered by the two frequencies previously selected.
		
		\subsubsection*{Method \#2}
\begin{figure}
			\centering
			\includegraphics[trim=0 0 0 0,width=\linewidth]{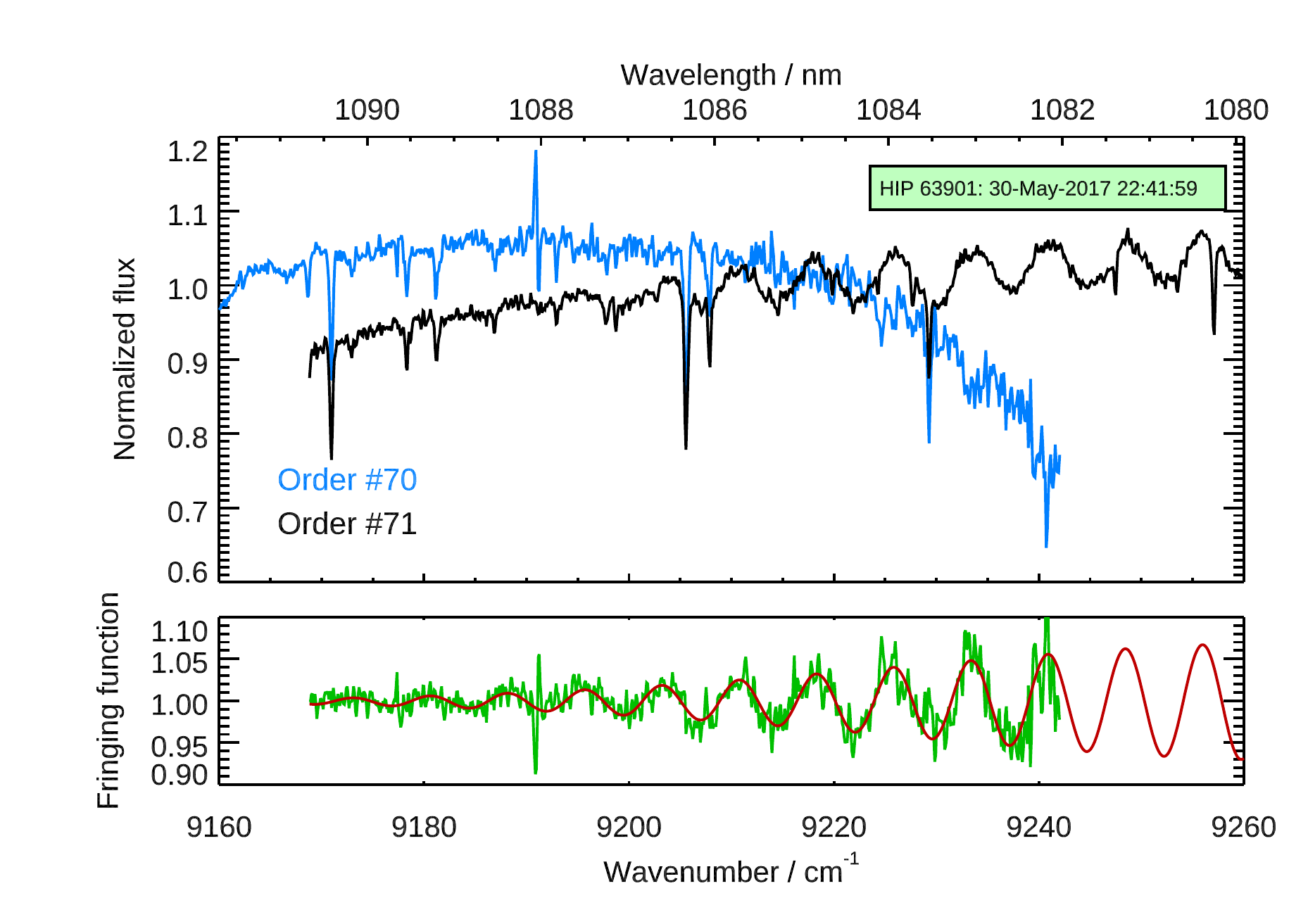}
			\caption{Example of fringing pattern in the spectrum of a telluric standard.  Top panel: spectral orders \#70 and \#71.  Bottom panel: ratio of the two spectral orders after removal the ratio of the blazing functions (green) and its fit (red).}
			\label{fringing3}
		\end{figure}

\begin{figure*}
			\centering
			\includegraphics[width=17cm]{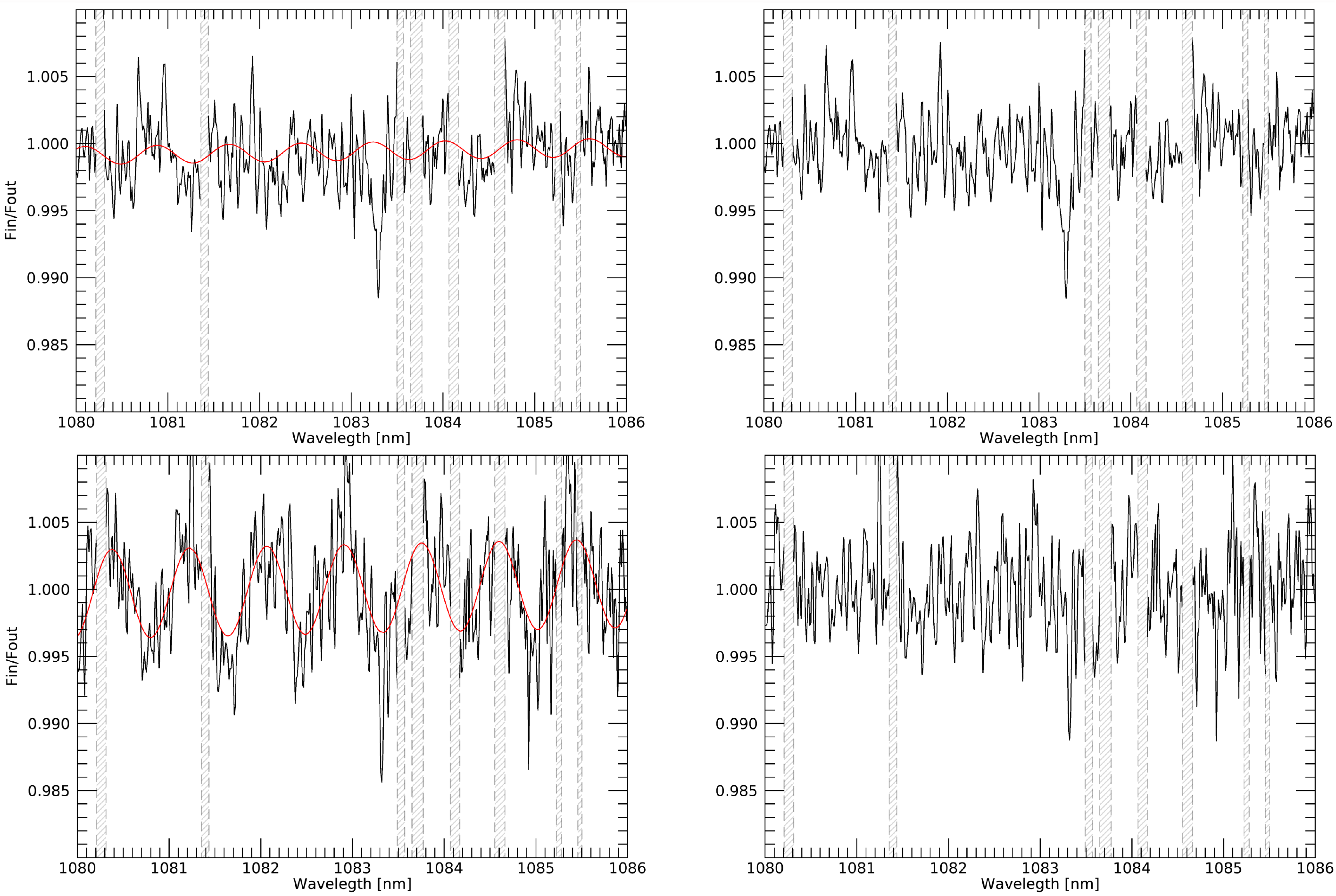}
			\caption{Fringing correction for the transit 3 (upper panels) and 1 (lower panels). The mean in-transit transmission spectra in the stellar rest frame for the A-nodding position before the fringing correction, with overplotted $F_{frin}$ (in red), are shown in the left panels, while the right panels show the fringing corrected mean spectra. The telluric line residuals are masked (in gray).}
			\label{fringing}
		\end{figure*}

		The third method exploits the very different fringing pattern amplitudes in overlapping regions of adjacent spectral orders on the detector.  In particular, while the signal-to-noise ratio is highest near the \ion{He}{I}~1083.3~nm line recorded at spectral order \#71, the fringing pattern is practically absent at the wavelengths recorded at order \#70 (Figure~\ref{fringing3}, top panel).
		
		If we divide the higher signal-to-noise ratio spectrum seen at order \#71 by the same spectrum recorded at order \#70, both the stellar and the telluric spectrum are removed.  Once the ratio of the blaze functions of the two orders is also removed, only the fringing pattern at order \#71 is left.  Since the amplitude of the pattern varies noticeably in the overlapping range of the two spectral orders ($\sim 1082 - 1090$~nm), we fit this pattern in wavenumber as shown in the bottom panel of Fig.~\ref{fringing3} using a sine function folded with a Gaussian centred at the peak of the blaze function, $k_0$ (fixed to the value 9267.84~cm$^{-1}$):
		\begin{displaymath}
			A\:e^{-B\:(k-k_0)^2}\:\sin\left[2\pi(k-k_1)/K\right] + C,
		\end{displaymath}
		where the free parameters of the fit function are: $K$ (the frequency of the pattern), $A$ (the amplitude), $B$ (the scale of attenuation), $k_1$ (the phase), and $C$ (the overall offset).
		
		It is worth noting that the fringing function is remarkably stable among all the data sets we have considered, even though there are some differences among the different transit dates in phase or in amplitude. We also verify that the sine-wave function is a reasonably good approximation in the middle of the overlapping range (1084-1087~nm).  The average pattern frequency, in particular, turns out to be $7.66 \pm 0.07$~ cm$^{-1}$.
		
		\subsubsection*{Comparison between methods}
		
\begin{figure*}
			\centering
			\includegraphics[width=\linewidth]{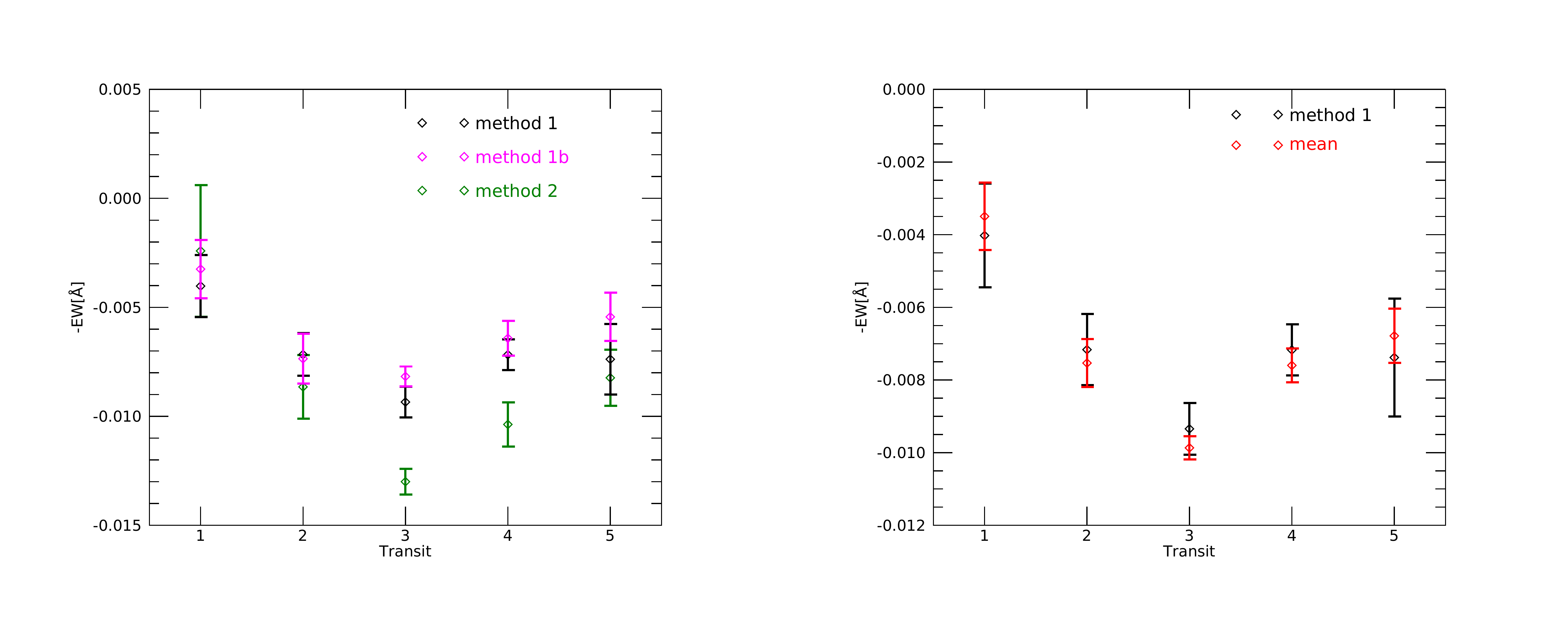}
			\caption{Comparison between the three fringing correction methods. The EWs$_{\rm AB}$ values for the five nights (x-axis) averaged between the contact point  $t_2$ and $t_3$ are plotted. Left panel the results of the three methods (method \#1 in black, method \#1b in magenta, and method \#2 in green).  Right panel:  comparison between method \#1 and the weighted average of the three methods.}
			\label{fringing_2}
		\end{figure*}

		As the left panel of Fig.~\ref{fringing_2} shows, the absorption values obtained at the \ion{He}{I} position during the 5 planetary transits are consistent using these three different techniques.  For this reason we are confident with our fringing removal methods. In particular, when performing the data analysis on the transmission spectra during transit, we adopt correction Method \#1. Indeed, as shown in the right panel of Fig.~\ref{fringing_2}, the results from this Method are very close to the weighted average of the three methods. Method \#2 is adopted for the out-of-transit stellar variability study.
		
		
		\section{In-transit H$\alpha$ vs \ion{Na}{I} vs \ion{He}{I} relation} \label{App_B}
\begin{figure}
			\includegraphics[width=9cm]{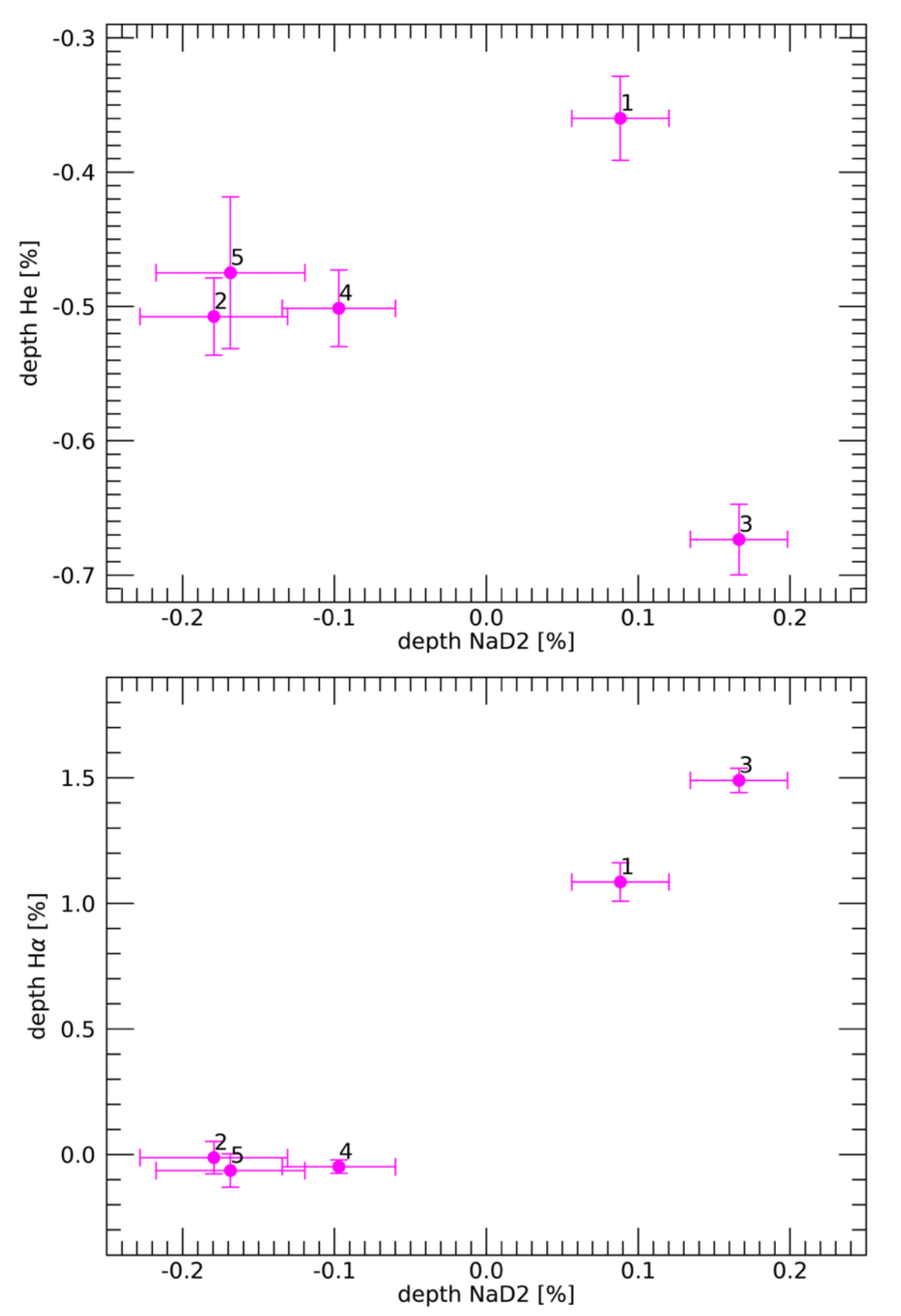}
			\caption{In-transit H$\alpha$ vs. Na D$_2$ vs. \ion{He}{I} relation. The He vs. Na D$_2$ transit depths (upper panel) follow a similar behaviour to that highlighted by the \ion{He}{I} vs. H$\alpha$ relation (Fig.~\ref{He_ha}). This is a consequence of the direct proportionality between Na D$_2$ and  H$\alpha$ proxies (lower panel).}
			\label{na}
		\end{figure}

		As a consistency check of the analysis presented in \S~\ref{He_Ha_relatio}, we also investigate the behaviour of the stellar \ion{Na}{I} D lines. In particular, we focus on the D$_2$~588.9951~nm line as it has a higher line contrast compared to the D$_1$~589.5924~nm line. We use the same light-curve-based approach followed for H$\alpha$ in \S~\ref{He_Ha_relatio}. A plot of the \ion{He}{I} vs. \ion{Na}{I} D$_2$ transit depths highlights an almost identical behaviour with respect to the \ion{He}{I} vs. H$\alpha$ relation, stemming from the direct proportionality between the H$\alpha$ and \ion{Na}{I} D$_2$ signals (see upper and lower panels of Fig.~\ref{na}, respectively). 
		This confirms the expectations for similar in-transit behaviour of H$\alpha$ and \ion{Na}{I} lines from the analysis of  \S~\ref{stellarvariability} and agrees with what is expected from modeling efforts \citet{Cauley2018}, thereby reinforcing the reliability of the results presented in \S~\ref{He_Ha_relatio}.

	\end{appendix}
\end{document}